\documentclass[11pt,a4paper]{article}
\usepackage{jcappub}

\usepackage{bbm,bm}
\usepackage{xcolor}
\usepackage{subcaption}
\usepackage{tikz}
\usepackage{cancel}
\usepackage{empheq}
\allowdisplaybreaks[1]
\newcommand{\la}[1]{\label{#1}}
\newcommand{\be}{\begin{equation}}
\newcommand{\ee}{\end{equation}}
\newcommand{\ba}{\begin{eqnarray}}
\newcommand{\ea}{\end{eqnarray}}
\newcommand{\rmi}[1]{{\mbox{\scriptsize #1}}}
\newcommand{\fig}{fig.~}
\newcommand{\figs}{figs.~}
\newcommand{\eq}{eq.~}
\newcommand{\eqs}{eqs.~}
\newcommand{\se}{sec.~}

\newcommand{\app}{appendix~}

\newcommand{\nr}[1]{(\ref{#1})}
\newcommand{\tr}{{\rm Tr\,}}
\newcommand{\nn}{\nonumber \\}

\def\undertilde#1{\mathop{\vtop{\ialign{##\cr$\textstyle{#1}$\cr%
\noalign{\kern1pt\nointerlineskip}\hfil$\mathchar"0365$\hfil\cr}}}}
\def\wideundertilde#1{\mathop{\vtop{\ialign{##\cr$\textstyle{#1}$\cr%
\noalign{\kern1pt\nointerlineskip}\hfil$\mathchar"0367$\hfil\cr}}}}

\renewcommand{\vec}[1]{{\bf #1}}

\newcommand{\Nc}{N_{\rm c}}

\newcommand{\F}{\mathcal{F}}

\renewcommand{\P}{\mathcal{P}}

\newcommand{\dA}{d_\rmii{A}}

\newcommand{\rmO}{{\mathcal{O}}}
\newcommand{\bmu}{\bar\mu}

\def\lsi{\raise0.3ex\hbox{$<$\kern-0.75em\raise-1.1ex\hbox{$\sim$}}}
\def\gsi{\raise0.3ex\hbox{$>$\kern-0.75em\raise-1.1ex\hbox{$\sim$}}}

\newcommand{\rmii}[1]{{\mbox{\tiny\rm{#1}}}}

\newcommand{\Tint}[1]{{\hbox{$\sum$}\!\!\!\!\!\!\!\int\,}_{\!\!\!\!\raise-0.9ex\hbox{$\scriptstyle{#1}$}}}
\newcommand{\Tinti}[1]{{{\Sigma}\!\!\!\!\raise0.3ex\hbox{$\int$}_\rmii{${#1}$}}}
\newcommand{\Tintip}[1]{{{\Sigma'}\!\!\!\!\!\raise0.3ex\hbox{$\int$}_\rmii{${#1}$}}}
\newcommand{\bi}{\begin{itemize}}
\newcommand{\ei}{\end{itemize}}
\newcommand{\hide}[1]{ }

\newcommand{\naive}[2]{\widehat{#1}^{ }_{#2}}
\newcommand{\scale}{\ell}

\newcommand{\rev}[1]{#1} 

\makeatletter \@addtoreset{equation}{section} \makeatother
\renewcommand{\theequation}{\arabic{section}.\arabic{equation}}
\makeatletter
\renewcommand\section{\@startsection {section}{1}{\z@}%
                                   {-5.5ex \@plus -1ex \@minus -.2ex}
                                   {2.3ex \@plus.2ex}%
                                   {\normalfont\large\bfseries}}
\renewcommand\subsection{\@startsection{subsection}{2}{\z@}%
                                     {-3.25ex\@plus -1ex \@minus -.2ex}%
                                     {1.5ex \@plus .2ex}%
                                     {\normalfont\normalsize\bfseries}}
\renewcommand\thesection {\@arabic\c@section}
\renewcommand\thesubsection   {\thesection.\@arabic\c@subsection}
\renewcommand{\@seccntformat}[1]{%
\csname the#1\endcsname.\hspace{1.0em}}
\makeatother

\usetikzlibrary{arrows.meta}
\newcommand{\myarrow}[6]{
\draw[thick, line width=0.75mm, #6, >={Latex[length=2mm, width=2mm]},
  color=#5] (#1, #2) -- (#3, #4);
}
\newcommand{\plaquette}[5]{
\begin{tikzpicture}
    \myarrow{0}{0}{1}{0}{#1}{#5};
    \myarrow{1}{0}{1}{1}{#2}{#5};
    \myarrow{1}{1}{0}{1}{#3}{#5};
    \myarrow{0}{1}{0}{0}{#4}{#5};
\end{tikzpicture}
}

\newcommand{\clover}{
\begin{tikzpicture}
    \node at (0,0){\plaquette{blue}{blue}{purple}{purple}{->}};
    \node at (1.2,0){\plaquette{blue}{blue}{purple}{purple}{->}};
    \node at (0,1.2){\plaquette{blue}{blue}{purple}{purple}{->}};
    \node at (1.2,1.2){\plaquette{blue}{blue}{purple}{purple}{->}};
\end{tikzpicture}
}
\newcommand{\clovercc}{
\begin{tikzpicture}
    \node at (0,0){\plaquette{purple}{purple}{blue}{blue}{<-}};
    \node at (1.2,0){\plaquette{purple}{purple}{blue}{blue}{<-}};
    \node at (0,1.2){\plaquette{purple}{purple}{blue}{blue}{<-}};
    \node at (1.2,1.2){\plaquette{purple}{purple}{blue}{blue}{<-}};
\end{tikzpicture}
}

\begin{document}

\flushbottom




\begin{flushright}
April 2026
\end{flushright}

\vspace*{-0.9cm}

\arxivnumber{2601.09784}


\title{\boldmath\Large
 Classical equipartition dynamics between \\[0mm] 
 axions and non-Abelian gauge fields
}


\author[a]{\large Kim V.\ Berghaus,} 
\author[b]{\large Adrien Florio,} 
\author[c]{\large M.~Laine,} 
\author[b]{\large Franz R.\ Sattler\,} 


\affiliation[a]{
Walter Burke Institute for Theoretical Physics, \\
California Institute of Technology, CA 91125, USA
}


\affiliation[b]{
Fakult\"at f\"ur Physik, Universit\"at Bielefeld, 
D-33615 Bielefeld, Germany
}


\affiliation[c]{
AEC, 
Institute for Theoretical Physics, 
University of Bern, \\ 
Sidlerstrasse 5, CH-3012 Bern, Switzerland
}

\emailAdd{berghaus@caltech.edu}
\emailAdd{aflorio@physik.uni-bielefeld.de}
\emailAdd{laine@itp.unibe.ch}
\emailAdd{fsattler@physik.uni-bielefeld.de}


 

 
%
\abstract{
Motivated by axion-like inflation and its warm embedding
within the Standard Model, we study the early stages of the energy
transfer between an axion condensate and an SU(2) gauge ensemble, 
by employing non-linear classical real-time lattice simulations. 
The discretized equations of motion are worked out, 
elaborating on Gauss constraints.  
A numerical solution is implemented on the  
\href{https://github.com/cosmolattice/cosmolattice}{CosmoLattice}
platform. Adopting a quadratic potential, 
and omitting universe expansion, 
we establish initial 
exponential growth of the low-momentum gauge modes;
damping of axion oscillations after some delay; 
and subsequent energy equipartition between axion and gauge ensembles. 
A clear difference between 
the SU(2) and U(1) dynamics is observed, likely associated with non-Abelian
self-interactions. We elaborate on what this implies for the possible
thermalization of the SU(2) ensemble.  
}


 

\maketitle

%
\section{Introduction}
\la{se:intro}

Steady improvements in cosmological data~\cite{planck6,act,spt} 
are making inflationary model building ever more interesting. 
{}From the theoretical perspective, one particularly 
nice class of models falls under the name of natural 
inflation~\cite{ai}, where the desired flatness of the 
inflationary potential is guaranteed by a shift symmetry.  
The inflaton is then typically a pseudoscalar field, 
reminiscent of the QCD axion, originally proposed to exist 
for other reasons~\cite{cp1,cp2,cp3}. 
Though explicit constructions 
are not quite easy to formulate
(cf., e.g., ref.~\cite{heavy_ax} for the past status), 
this general class of models has become rather popular
in recent years.

An axion-like field couples to 
the topological charge density, $F\widetilde F$, 
of all available gauge fields. Much work has been 
devoted to the coupling to Abelian fields, as the
associated tachyonic instability may lead
to spectacular signatures~\cite{as,as2}.
In the meanwhile, these studies have reached 
an advanced level, with numerical
simulations offering a tool for
fully capturing the non-linear dynamics
(cf., e.g., refs.~\cite{axab_0,axab_1,axab_2,axab_3,axab_4} 
and references therein).

However, the dynamics experienced by non-Abelian
fields could differ from the Abelian case. 
Non-Abelian gauge covariance induces momentum-dependent cubic
self-interactions, which drive rapid equilibration~\cite{equil}.
Though a large Hubble rate poses a challenge for this, it is still interesting 
to ask what happened if equilibration were reached 
(cf., e.g., refs.~\cite{fixed_pt,therm,hook,ema}).  
It turns out that the so-called sphaleron processes of the 
non-Abelian plasma generate a thermal friction for the slowly
evolving axion condensate~\cite{mms,warm}. 
Estimates of the sphaleron rate in 
a weakly coupled SU(3) plasma have been obtained 
with classical real-time lattice simulations~\cite{mt,clgt}.
Making use of them, a scenario of  
warm inflation can be implemented~\cite{mwi}, with 
the regular objections~\cite{linde} avoided by
the approximate shift symmetry. 
In particular, 
given an improved understanding of warm-inflation curvature
power spectra~\cite{mehrdad,ballesteros,freese,ramos,sm3}, 
it can be shown that successful warm inflation can be achieved
with the QCD axion coupling~\cite{sm1} (see also refs.~\cite{sm2,sm3}), 
utilizing a quartic UV potential that would be excluded without
the help of the thermal friction. This class of models 
makes characteristic predictions about other standard
observables, like non-Gaussianities~\cite{Bastero-Gil:2014raa, mehrdad, mm}, 
gravitational waves~\cite{sorbo,reheat,klose,new,new2}, 
or maximal post-inflationary temperatures~\cite{helena}.

We may wonder to what extent  
\rev{the main qualitative 
features of the non-Abelian \linebreak
sphaleron heating mechanism}
depend on whether the system actually equilibrates.  
Like in the Abelian case with the tachyonic instability,  
some energy transfer certainly takes place between the
axion condensate and long-wavelength gauge fields. In a non-Abelian
situation, self-interactions due to cubic and quartic gauge  
vertices should efficiently redistribute the energy insertions.
As the instability band lies at small momenta, the redistributed
energy will also initially be in the long-wavelength modes. 
It is precisely
the long-wavelength gauge fields that are responsible for the 
sphaleron processes, so we might expect that
an analogue thereof could also exist in 
a non-equilibrium setting.

The purpose of our study is to  
simulate the non-equilibrium dynamics numerically. 
To show how the relevant observables can be extracted, we first do 
this in Minkowskian spacetime, where Hubble damping does not 
complicate the interpretation. We span the regime from underdamped 
to overdamped oscillations, with the former giving 
some insight into early matter domination dynamics, 
and the latter resembling more closely 
what happens during inflation. In particular, 
from the overdamped regime we can draw first parallels 
to warm inflation, which we hope to consolidate 
in a future study that includes Hubble expansion. 

Given that we are not aware of previous numerical simulations 
of the very same framework, we also invest some effort in spelling out 
its conceptual 
and technical foundations. 
That said, the non-Abelian setup has been studied 
perturbatively in the inflationary context
(cf., e.g., refs.~\cite{Domcke:2019lxq,pencil}), and there are also 
numerical investigations which did include a non-Abelian
$F\widetilde F$-operator
(cf., e.g., ref.~\cite{bau} and references therein),
though in the latter case as a small correction to dynamics
that was dominated by renormalizable gauge interactions. 
In a separate line of research, 
a gauge-singlet inflaton field coupling to a non-Abelian
$F F$ rather than $F\widetilde F$-operator
has been studied~\cite{scalar}.

Our presentation is organized as follows. 
We start by summarizing our setup in 
continuum notation (cf.\ \se\ref{se:summary}). 
Subsequently, some key features of the early-time solution are worked out
semi-analytically (cf.\ \se\ref{se:analytic}). 
We then present the implementation (cf.\ \se\ref{se:numerics}) and 
main findings (cf.\ \se\ref{se:results}) from 
our non-linear simulations. 
After drawing conclusions and offering an outlook 
(cf.\ \app\ref{se:concl}), 
we turn to a detailed specification of the equations
of motion and the associated conservation laws, first
in continuum (cf.\ \app\ref{se:cont}), 
and then in the presence of 
a spatial lattice discretization (cf.\ \app\ref{se:latt}).

%
\section{Summary of setup}
\la{se:summary}

We consider a system consisting of a pseudoscalar field, 
$\varphi \in \mathbbm{R}$,
and non-Abelian gauge fields. The $\varphi$ is assumed neutral with
respect to the gauge group. The Lagrangian density reads 
\be
 \mathcal{L} 
 \; = \;
     \frac{1}{2} \partial^\mu_{ }\varphi\, \partial^{ }_\mu \varphi
    - V(\varphi)
    - \frac{1}{4} F^a_{ \mu\nu}F^{a\mu\nu}_{ }
    - \frac{\varphi\,\chi}{f^{ }_a}
 \;, \la{L}
\ee
where $\chi$ is the topological charge density, 
\be
 \chi \; \equiv \; 
 \frac{ \epsilon^{\mu\nu\rho\sigma}_{ }
 g^2 F^{a}_{\mu\nu} F^{a}_{\rho\sigma} }{64\pi^2_{ }} 
 \;. \la{chi}
\ee
Here, we have assumed the metric signature ($+$$-$$-$$-$);
$F^{a}_{ \mu\nu}$ is the Yang-Mills field strength, 
with $a = 1, ..., \Nc^2 - 1$ labelling the generators; 
$\epsilon^{0123}_{ } = 1$; 
and $g^2_{ } = 4\pi\alpha$ is the gauge coupling. 
To reduce the numerical cost, 
we focus on $\Nc^{ } = 2$ in the present study, 
even though for warm inflation with the QCD axion, 
$\Nc^{ } = 3$ would be the relevant case. 
In order to simplify the equations, we parametrize the 
coupling between $\varphi$ and the gauge fields via
\be
 \kappa \; \equiv \; \frac{8 g^2_{ }}{64\pi^2_{ }f^{ }_a}
 \; \equiv \; \frac{\alpha}{2\pi f^{ }_a}
 \;, \la{kappa}
\ee 
such that the axion--gauge interaction reads 
\be
 \mathcal{L} 
 \;\supset\;
 - \frac{\kappa \varphi}{8} 
 \epsilon^{\mu\nu\rho\sigma}_{ }
 F^{a}_{\mu\nu} F^{a}_{\rho\sigma}
 \; = \; 
 - \frac{\kappa \varphi}{2} 
 \epsilon^{0ijk}_{ }
 F^{a}_{0i} F^{a}_{jk}
 \; \overset{\rmii{\nr{F0i_full}}}{\equiv} \; 
    - \kappa\, \varphi\, F^a_{0i} B^a_i
 \, ,
\ee
where $ B^a_i \equiv \frac{1}{2}\epsilon^{ }_{ijk} F^a_{jk}$ 
denotes a non-Abelian magnetic field. 

The equations of motion following from the Lagrangian 
in \eq\nr{L} are derived in \app\ref{se:cont}.
Doing a partial fixing to the gauge $A^a_0 = 0$, and taking the spatial 
gauge potentials, $A^a_i$, as the canonical variables, 
the would-be Euler-Lagrange equation for
$A^a_0$ yields the {\em Gauss constraint} that needs to be 
satisfied at the initial time, 
\be
 \mathcal{D}\cdot \vec{E} 
 \;
 \overset{\rmii{\nr{Gauss_cont_pre}}}{=}
 \;
 \kappa\, (\nabla\varphi) \cdot \vec{B}
 \;, 
 \la{Gauss_cont}
\ee
where $\vec{E}$ and $\vec{B}$ are vectors both in gauge algebra 
and physical space, and $\mathcal{D}$ is the covariant derivative in the adjoint representation. The fact that $\vec{E}$ and $\vec{B}$ are 
defined in terms of gauge potentials, leads to two 
{\em Jacobi identities}, 
\be
 \mathcal{D}\cdot\vec{B}
 \; \overset{\rmii{\nr{jacobiI_pre}}}{=} \; 0 
 \vphantom{\Bigl| }
 \;, \quad 
 \dot{\vec{B}}
 \; \overset{\rmii{\nr{jacobiII_pre}}}{=} \;
 \mathcal{D}\times\vec{E} 
 \;. 
 \la{jacobiI}
\ee
Finally, 
the Euler-Lagrange equations for $\varphi$ and $A^a_i$, 
together with the definition of $E^a_i$ from 
\eq\nr{F0i_full}, 
yield the evolution equations for the dynamical variables, 
\begin{empheq}
      {align}
 \quad \vphantom{\Big | }
 \ddot{\varphi} & 
 \; \overset{\rmii{\nr{eom_cont_scalar}}}{=} \;  
 \nabla^2_{ }\varphi - V'(\varphi) - \kappa \,\vec{E}\cdot\vec{B}  
 \;, \la{eom_cont_scalar_0} \\[2mm]
 \dot{\vec{A}} & 
 \; \overset{\rmii{\nr{eom_cont_Ai}}}{=} \;
  \vec{E} 
 \;, \la{eom_cont_Ai_0} \\[2mm]
 \dot{\vec{E}} &
 \;
 \overset{\rmii{\nr{eom_cont_Ei}}}{=}
 \; 
 - 
  \mathcal{D}\times \vec{B}
 + \kappa 
 \bigl(\, 
   \dot\varphi\, \vec{B}
  - \nabla\varphi \times \vec{E}
 \,\bigr)
 \;. \quad \vphantom{\Big | }
 \la{eom_cont_Ei_0} 
\end{empheq}

To solve the evolution equations, we need to set initial 
conditions at time $t=0$. Motivated by the theory of cosmological inflation, 
we insert there a spatially constant condensate, 
${
 \varphi(t,\vec{x})\to \bar\varphi(t)
}$, 
which carries almost all of the original energy density. 
We choose a quadratic potential for the axion, 
\be
 V(\varphi) \; \equiv \; \frac{m^2_{ }\varphi^2_{ }}{2} 
 \;. \la{V}
\ee
For $\kappa = 0$,  
\eq\nr{eom_cont_scalar_0} then implies the presence of oscillations. 
It is convenient to start the evolution at a moment when 
the condensate crosses zero, so we impose
\be
 \bar\varphi(0) \; = \; 0
 \;, \quad
 \dot{\bar\varphi}(0) \; \neq \; 0
 \;. \la{initial_condensate}
\ee

Apart from the condensate, we also insert some noise at non-zero
momenta at $t=0$
(otherwise the gauge fields stay at zero).
It is conventional to 
mimick quantum fluctuations with such noise~\cite{cl1}. 
We thus write down an analogue of a mode expansion, 
\begin{empheq}
      {align}
 \quad
 \varphi(0,\vec x)
 & \; = \; 
 \int_\vec{k}^{ } 
 \Bigl[\,
   c^{ }_\vec{k} \, \varphi^{ }_k(0) \, e^{i \vec{k}\cdot \vec{x} }_{ }
  +
   c^*_\vec{k} \, \varphi^{*}_k(0) \, e^{-i \vec{k}\cdot \vec{x} }_{ }
 \,\Bigr]
 \;, \quad
 \int_\vec{k}^{ }
 \; \equiv \; \int^{ }_{ } 
 \! \frac{{\rm d}^3_{ }\vec{k}}{(2\pi)^3_{ }} 
 \;, 
 \la{initial_varphi}
 \\[3mm]
 \dot\varphi(0,\vec x)
 & \; = \; 
 \int_\vec{k}^{ }
 (i \omega^{ }_k)
 \Bigl[\,
   - c^{ }_\vec{k} \, \varphi^{ }_k(0) \, e^{i \vec{k}\cdot \vec{x} }_{ }
  +
     c^*_\vec{k} \, \varphi^{*}_k(0) \, e^{-i \vec{k}\cdot \vec{x} }_{ }
 \,\Bigr]
 \;, \quad
 \la{initial_dvarphi}
\end{empheq}
where $c^{ }_\vec{k}$ are complex random numbers, 
with $\langle c^{ }_\vec{k} \rangle \equiv 0$, and 
$\varphi^{ }_k(t)$ is a forward-propagating 
mode function, 
$
 \varphi^{ }_k(t) \equiv e^{-i \omega^{ }_k t}/\sqrt{2 \omega^{ }_k}
$
(cf.\ \eq\nr{canonical}). 

When modelling quantum fluctuations through classical fields, 
there is an issue of a factor 1/2 
(cf.\ the discussion below \eq\nr{def_P}). 
There are various possibilities for overcoming this, 
such as restricting to 1/2 of the full 
momentum space~\cite{cl1}; 
dividing the quantum mode
functions by $\sqrt{2}$~\cite{cl2}; 
or simply inserting
an extra factor in the noise variance, 
\be
 \langle\, c^{ }_\vec{k} \, c^{ }_\vec{q} \,\rangle
 \; = \; 0
 \;, \quad
 \langle\, c^{ }_\vec{k} \, c^{*}_\vec{q} \,\rangle 
 \; = \; 
 f^{ }_k\, 
 (2\pi)^3_{ }\delta^{(3)}_{ }(\vec{k-q})
 \;, \quad
 f^{ }_k 
 \; \equiv \; 
 \frac{\theta(\Lambda - k)}{2}
 \;. \la{variance}
\ee
We have used
$f^{ }_k$ to also impose an upper cutoff, $\Lambda$, 
on the momentum range, because
classical field theory suffers from a 
Rayleigh-Jeans divergence, and does not describe correctly
large momenta (physically, $k \ge T$). 
\rev{In the present study, we set} 
\be
 f^{ }_k \; \to \; 10^{-10}_{ } \, \theta(\Lambda - k)
 \;, \la{variance_new}
\ee
\rev{where $\Lambda$
is chosen much smaller than the inverse lattice spacing, in order to avoid
discretization artifacts. The very small overall noise amplitude is chosen
so that we can focus on classical field dynamics, without needing
to increase the initial condensate energy density beyond 
a numerically convenient magnitude.
}

For the gauge sector, we insert 
no background value in the initial state 
(however, nothing prohibits one from being generated dynamically under
the right conditions, as suggested in refs.~\cite{chromo1,chromo2}). 
We insert noise in the electric field, in analogy with
\eq\nr{initial_varphi}; the details 
are given in \eqs\nr{initial_Ei} [in continuum notation]
and \nr{initial_Ei_latt} [in lattice notation]. 
The initial magnetic fields, $B^a_i$, are set to zero, 
so as to satisfy the Gauss constraint from \eq\nr{Gauss_cont}. 
Non-zero $B^a_i$ arise rapidly from the dynamics, 
as implied by \eq\nr{eom_cont_Ai_0}. We remark that  
our qualitative results are stable in the class of initial conditions 
that satisfy the Gauss constraint.

\vspace*{3mm}

Among the most important observables that we measure are the
{\em power spectra} associated with different fields. Focussing
on the example of $\varphi$, we recall that 
in a translationally invariant system, 
statistical averages only depend on relative separations, 
$
 \langle \varphi(t,\vec{x}) \varphi(t,\vec{y}) \rangle
 = 
 \F(\vec{x-y})
 \;. 
$
For the corresponding Fourier transforms, 
$
 \varphi(t,\vec{k}) \equiv \int_\vec{x}
 e^{-i\vec{k}\cdot\vec{x}}_{ }
  \varphi(t,\vec{x})
$, 
where
$
 \int_\vec{x} \equiv \int \! {\rm d}^3_{ }\vec{x}
$,
this implies that 
\ba
 \langle \varphi(t,\vec{k}) \varphi^*_{ }(t,\vec{q}) \rangle
 & = & 
 (2\pi)^3_{ }\delta^{(3)}_{ }(\vec{k-q})
 \, G^{ }_\varphi(t,\vec{k})
 \;, \\[2mm]
 G^{ }_\varphi(t,\vec{k})
 & \equiv & 
 \int_\vec{x} 
 e^{-i\vec{k}\cdot\vec{x}}_{ }
 \langle \varphi(t,\vec{x}) \varphi(t,\vec{0}) \rangle
 \;. \la{P_cl_pre}
\ea
Assuming that the Fourier transform is independent of the 
direction of $\vec{k}$, the actual power spectrum is  
defined by multiplying $G^{ }_\varphi$ with the phase-space volume element, 
\be
 \P^{ }_{\varphi}(t,k)
 \; \equiv \; 
 \frac{k^3_{ } G^{ }_\varphi(t,\vec{k})}{2\pi^2_{ }} 
 \;. \la{P_cl}
\ee
The electric and magnetic power spectra, 
$
 \P^{ }_{E}
$
and
$
 \P^{ }_{B}
$, 
are defined analogously. We remark that in the non-Abelian
case, the power spectra $
 \P^{ }_{E}
$
and
$
 \P^{ }_{B}
$ 
are {\em not} gauge invariant as functions of $k$, 
as they involve
correlating two different spatial positions (cf.\ \eq\nr{P_cl_pre}).
However, integrals over the power spectra are so, as they correspond
to an equal-position correlator, 
\be
 \langle\, \varphi(t,\vec{x}) \varphi(t,\vec{x}) \,\rangle
 \; = \; 
 \int_{0}^{\infty} \! \frac{ {\rm d} k}{k} \, 
 \P^{ }_{\varphi}(t,k)
 \;. 
\ee 

\vspace*{3mm}

To conclude this section, we note that all the fields in our
system, and some of the parameters ($\kappa$, $m$), are dimensionful. 
In a numerical study, we need to express dimensionful quantities in 
units of some arbitrary  scale. 
We choose for this role a length
scale, denoted by~$\scale$. The scale should be set so that 
dimensionful parameters, and also initial
conditions (cf.\ \eq\nr{initial_condensate}), are of order unity, 
\be
 \widetilde\kappa 
 \; 
 \overset{\rmii{\nr{kappa}}}{\equiv} 
 \; \frac{\kappa}{\scale} 
 \; \sim \; 1
 \;, \quad
 \widetilde m 
 \; 
 \overset{\rmii{\nr{V}}}{\equiv} 
 \; \scale \, m 
 \; \equiv \; 1
 \;, \quad
 \scale^2_{ }\dot{\bar\varphi}(0)
 \; 
 = 
 \; 
 \partial^{ }_{\tilde t}\;\bar{\!\widetilde\varphi}(0)
 \; 
 \overset{\rmii{\nr{initial_condensate}}}{\sim} 
 \; 
  1
 \;, \la{rescale0}
\ee
where the middle relation shows that in practice we have 
fixed $\scale$ in terms of the axion mass. 
\rev{We employ a generic $\ell$ for the scale setting,
rather than for instance the axion mass, in order to ensure the
model independence of our framework.}
Time and momentum are also rescaled, 
$\tilde t \equiv t / \scale$ and $\tilde k \equiv \scale\hspace*{0.3mm} k$.
The same rescalings are applied on the lattice 
(cf.\ \eqs\nr{rescale1}--\nr{rescale3}). 
Keeping the lattice spacing small in these units, 
$\tilde a \equiv a/\scale \ll 1$, 
guarantees the absence of discretization artifacts.

%
\section{Perturbative expectations}
\la{se:analytic}

The system introduced in \se\ref{se:summary}
displays some well-known remarkable properties. 
Here, we recall the origin of the tachyonic instability, 
in order to compare it later with non-linear simulations
(cf.\ \se\ref{ss:equi}). 

Let us consider a spatially homogeneous background value of 
$\varphi$, denoted by $\varphi(t,\vec{x}) \to \bar\varphi(t)$,
like in \eq\nr{initial_condensate}.
We want to derive an equation of motion for $\vec{B}$ in its
presence, staying at linear order. Taking the curl of \eq\nr{eom_cont_Ei_0}, 
and making use of the Jacobi identities 
in \eq\nr{jacobiI}, yields
\be
 \ddot{\vec{B}} 
 \; 
 \underset{\rmii{\nr{jacobiI}}}{
 \overset{\nabla\;\times\;\rmii{\nr{eom_cont_Ei_0}}}{ \approx }}
 \; 
 \nabla^2_{ }\vec{B}
  + \kappa \dot{\bar\varphi}\,\nabla\times\vec{B} 
 \;. \la{eom_B} 
\ee
Going to momentum space in the spatial directions, 
and solving for the eigenvalues, 
the circular polarization states ($\pm$) evolve as 
\be
 \ddot{B}^{ }_{k,\pm}
 \; 
 \overset{\rmii{\nr{eom_B}}}{\approx}
 \; 
 -\,
 k\, \bigl( \, k \pm \kappa \dot{\bar\varphi} \,\bigr) \, B^{ }_{k,\pm}
 \;. \la{B_modes}
\ee
When $k < \kappa |\dot{\bar\varphi}|^{ }_\rmi{max}$, there is the 
possibility of exponential growth. 

In order to solve \eq\nr{B_modes}, we need initial conditions. 
Though we had set the initial magnetic field to vanish, a non-zero value
is generated very rapidly through
\eq\nr{eom_cont_Ai_0}, so in practice we can take a configuration
carrying the same amount of energy density as the electric field 
in \eq\nr{initial_Ei}, as given by \eq\nr{initial_Bi}. 
The overall amplitude squared of the initial state
is given by \eq\nr{variance_new}. 
We let it phase-rotate according to the massless 
mode function, 
\be
 B^{ }_{k,\pm}(0) 
 \; = \; \sqrt{ k f^{ }_k }
 \;, \quad
 \dot{B}^{ }_{k,\pm}(0) 
 \; = \; 
 -\, i k B^{ }_{k,\pm}(0)
 \;. \la{B_initial}
\ee
After rescaling to a dimensionless quantity $\tilde{B}$ 
like in \eq\nr{rescale0}, 
we plot the power spectrum  
\be
 \P^{ }_{\tilde B}
 \; 
 \overset{\rmii{\nr{P_cl}}}{\underset{\rmii{\nr{rescale1}}}{\equiv}}
 \; 
 \frac{ 3 \tilde k^3_{ } }{2 \pi^2_{ }}
 \bigl[\, 
   |\widetilde B^{ }_{k,+}(\tilde t)|^2_{ }
  + 
   |\widetilde B^{ }_{k,-}(\tilde t)|^2_{ }
 \,\bigr]
 \;, \la{B_P}
\ee
where the factor 3 counts the SU(2) generators. 

\begin{figure}[t]

\hspace*{-0.1cm}
\centerline{%
 \includegraphics[width=0.48\linewidth]%
  {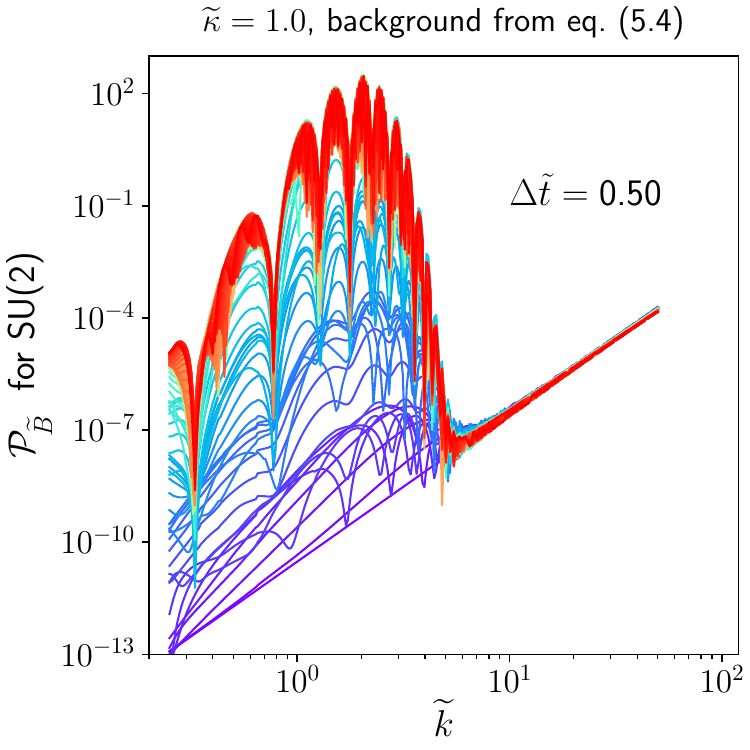}%
 \hspace*{0.2cm}
 \includegraphics[width=0.48\linewidth]%
  {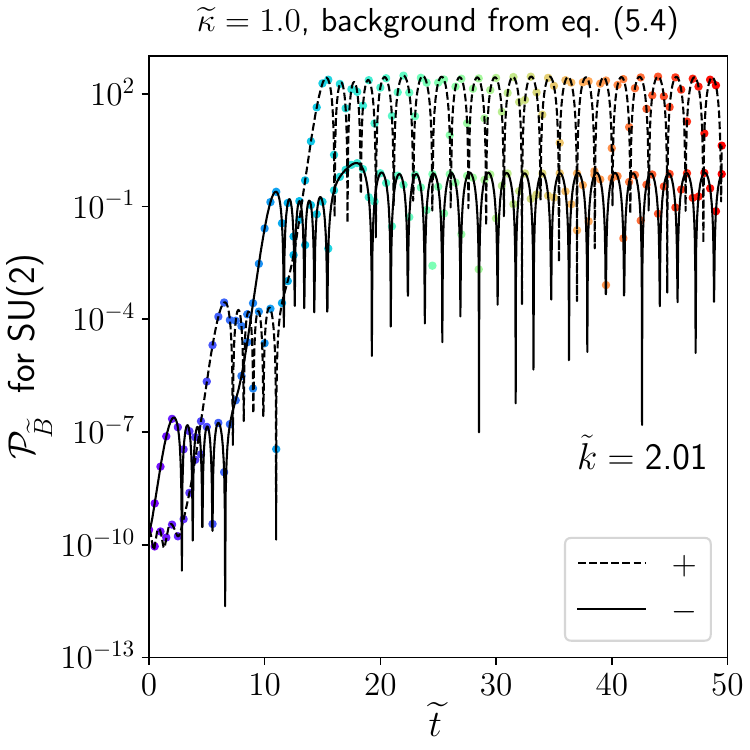}%
}

\caption[a]{\small
  Left: A numerical solution of \eq\nr{B_modes}, 
  with initial conditions from \eq\nr{B_initial},
  background modelling from \eq\nr{ansatz}, 
  and the 
  result plotted in terms of the power spectrum from 
  \eq\nr{B_P}. The curves correspond to snapshots
  at equal time intervals, $\Delta \tilde t = 0.50$,  
  and they display exponential growth for as long as 
  $\dot{\bar{\varphi}}\neq 0$ in \eq\nr{B_modes}. As a function
  of~$\tilde k$, they also show a band structure, characteristic
  of systems with periodically varying coefficients. The overall
  time period is the same as in \fig\ref{fig:power_spectra}, however the 
  linearized solution lacks the transfer of power to higher momenta
  that is seen there. 
  Right: The time evolution of the positive and negative
  helicity modes, at fixed $\tilde k = 2.01$. The 
  blobs correspond to the snapshots shown in the left panel
  (there both helicities are summed together).
  }
\la{fig:modes}
\end{figure}

\rev{
For the condensate, we assume a form justified
by our SU(2) simulation data, 
\begin{equation}
	\bar{\!\widetilde\varphi}(\,\tilde t\,) 
	 \; \approx  \;
	 \begin{cases}
	 	 \mathcal{A}_0^{ }\sin(\widetilde\omega_1^{ } \tilde t \,)\,, 
         & \text{for }\tilde t < \tilde t_0^{ } \,, \\[2mm]
	 	\sin(\widetilde\omega_2^{ } \tilde t - \phi_2^{ })\,
        \left\{\exp\left(-\widetilde\Upsilon(\tilde t - \tilde t^{ }_0)\right)
        (\mathcal{A}_0^{ }-\mathcal{A})+\mathcal{A}\right\}
	 	\,, & \text{for }\tilde t \geq \tilde t_0^{ }\,, 
	 \end{cases}
	 \label{ansatz} 
\end{equation}
whose interpretation is discussed in more detail
in a paragraph following \eq\nr{def_bar_varphi}.}
The coefficient
$\widetilde\Upsilon > 0$ accounts for the fact that oscillations
decay after an unattenuated early time period, 
of duration~$\tilde t^{ }_0$. 
After fixing the parameters from 
a comparison with lattice data 
(cf.\ \figs\ref{fig:oscillations}--\ref{fig:Upsilon}),  
we show snapshots of the solution for $\P^{ }_{\widetilde B}$ 
in \fig\ref{fig:modes}, taken at equal time intervals. For
$\tilde k < \widetilde\kappa\, |\dot{\bar{\widetilde \varphi}}|_{\text{max}}$, 
there are bands of exponentially growing solutions. In the non-linear 
dynamics, we will see later on 
that the band structure gets smoothened 
by the interactions
(cf.\ \fig\ref{fig:power_spectra}),  
but exponential growth persists. It depletes the energy 
density in the axion condensate, 
and ultimately leads to the equipartition
of the axion and gauge field 
energy densities (cf.\ \fig\ref{fig:equi}).

%
\section{Numerical implementation}
\la{se:numerics}

We now summarize the discretized
version of \se\ref{se:summary}, 
with the details relegated to \app\ref{se:latt}. The numerics has been 
implemented on the 
\href{https://github.com/cosmolattice/cosmolattice}{CosmoLattice}
platform~\cite{cosmo1,cosmo2}. 

%
\subsection{Spatial discretization}
\la{ss:finite_a}

In order to study the dynamics numerically, 
we discretize the spatial directions with a lattice spacing, 
$a > 0$. While doing this, it is important to maintain gauge 
covariance. We achieve this by following the strategy of standard 
lattice gauge theories, in which the spatial gauge potentials, $A^a_i$, 
are replaced by unitary link matrices, $U^{ }_i \in\,$SU(2), 
as the dynamical variables. 
The corresponding equations of motion are shown 
in \eqs\nr{eom_latt_scalar_resc}--\nr{eom_Ei_resc}.
As demonstrated in \app\ref{ss:energy}, the time evolution of
the spatially discretized equations conserves
the total energy of the system exactly, 
as long as time is a continuous coordinate. Comparing 
with the physical length scale, $\scale$ (cf.\ \eq\nr{rescale0}), 
our default lattice spacing is 
\be
 \tilde a \; \equiv \; \frac{a}{\scale}
 \; \approx \; 0.0332 \; \ll \; 1.0
 \;, \la{def_tilde_a}
\ee
\rev{
whose small value guarantees the absence of discretization
effects.}

%
\subsection{Time discretization and finite-volume effects}
\la{ss:time_vol}

In a simulation, the setup in 
\eqs\nr{eom_latt_scalar_resc}--\nr{eom_Ei_resc} needs 
to be approximated in two further ways: by discretizing
the time coordinate, 
and by placing the system in a finite volume. 

As for the time coordinate, we choose a step 
$\delta \tilde t \ll \tilde a$, and solve the equations with 
a 3rd order Runge-Kutta algorithm~\cite{rk1,rk2}. This algorithm is
{\em not} symplectic, meaning that energy is not conserved exactly. 
However, by restricting to  
\be 
 \delta \tilde t \; \le \; 0.15 \, \tilde a
 \;, 
\ee 
we find that energy is conserved up to relative accuracy $10^{-4}_{ }$ 
until times $\tilde t \le 50$ for $\tilde\kappa \approx 2.0$, 
which is sufficient for our purposes. 
For smaller values of~$\tilde\kappa$, a more accurate energy conservation
is observed, and we can integrate up to later times.  

As for the spatial volume, we adopt a cubic box of size $N^3_{ }$, 
with periodic boundary conditions in all directions. 
Tests have been carried out with
$ N = 256, 512, 1024$, verifying the absence of significant volume
dependence. Our SU(2) production runs, whose results are reported in 
\se\ref{se:results}, 
have been obtained with a fixed volume
\be
 N^3_{ } \; = \; 756^3_{ }
 \;. \la{def_N}
\ee
The U(1) production runs have identical parameters, except that
the default volume is $N^3_{ } = 512^3_{ }$. However, we have verified
the volume-independence of our results at $N \ge 256$.

%
\subsection{Initial conditions and power spectra}
\la{ss:initial}

While we keep the initial conditions for the condensate fixed, 
cf.\ \eq\nr{initial_condensate}, and make sure that it carries
most of the initial energy density, we have tested various 
recipes for supplementing the condensate with noise. 
In the default setup, 
the noise for $\widetilde\varphi$, 
which in continuum is given by 
\eqs\nr{initial_varphi}--\nr{variance_new},
is implemented according to 
\eqs\nr{initial_varphi_lat} and \nr{initial_dvarphi_lat}, 
respectively; the electric fields are initialized
according to \eq\nr{initial_Ei_latt}; and the magnetic
fields are set to zero, so as to satisfy the Gauss  
constraint from \eqs\nr{Gauss_kappa0} and \nr{Gauss_kappa1}.
In an alternative recipe, 
we put noise only in the magnetic fields 
(the links having been obtained from an exponential
map of gauge potentials), 
setting $\widetilde\varphi$ and $\widetilde E^a_i$ to zero. 
We have also 
tested cases where both electric and magnetic fields
are initialized simultaneously, though then the 
Gauss constraint from 
\eqs\nr{Gauss_kappa0} and \nr{Gauss_kappa1}
is not satisfied
(it would be 
customary to ``cool'' the fields, in order to bring
the violation of the Gauss constraint below a given 
tolerance, cf.,\ e.g.,\ refs.~\cite{cl2,old1,cl3}). While there are 
quantitative differences between these choices, 
moderately impacting the fit value of $\tilde{\Upsilon}$ 
in the parameterization of \eq\eqref{ansatz}, the
overall features of the equipartition dynamics stay the same.
For our SU(2) production runs, reported in 
\se\ref{se:results}, we use the default setup, 
with initial noise in $\widetilde\varphi$ and $\widetilde E^a_i$, 
but none in $\widetilde B^a_i$.

Once the evolution starts, 
equal-time power spectra are measured in accordance with 
\eqs\nr{P_cl_pre} and \nr{P_cl}, as a function of time. 
In lattice units, the power spectra are given 
by \eqs\nr{P_lat_pre} and \nr{P_lat}.

%
\section{Early stages of equilibration}
\la{se:results}

Given the setup outlined in \se\ref{se:numerics}, 
we proceed to the main results obtained
from our simulations. For non-Abelian dynamics we consider 
the example of an SU(2) Yang-Mills theory, and compare 
throughout with the U(1) theory. 
As for the parameters, we have fixed
\be
 \widetilde m 
 \; 
 \overset{\rmii{\nr{rescale0}}}{\equiv}
 \; 
 1
 \;, \quad 
 g 
 \; = \; 
 1.0
 \;, \quad
 \widetilde\kappa 
 \; = \; 
 0.2 \, ... \, 2.0
 \;, \quad
  \partial^{ }_{\tilde t} \; \bar{\!\widetilde\varphi}(0) 
 \; = \; 
 5.0
 \;. \la{params}
\ee
The value of $\widetilde m$ represents 
a choice of the scaling factor, $\scale$, and has no physical
significance. The largish gauge coupling, $g=1.0$, 
guarantees that non-Abelian
interactions are fairly efficient. The physically most 
interesting variation is in 
$\widetilde\kappa$, which parametrizes 
the strength of the interactions between the axion and 
gauge ensembles (cf.\ \eq\nr{kappa}).
\rev{The range indicated has been chosen
not for phenomenological or model building reasons, but in order to
interpolate between domains with qualitatively different dynamical
behaviours.}

The results that we show are in general based on a single
run in a large volume. When we take Fourier transforms, 
the smallest value shown 
is $\tilde k = 0.25$, corresponding 
to $\vec{n}=(1,0,0)$ 
in terms of \eq\nr{k_lat}. 
For larger momenta, 
a single lattice volume 
contains several wavelengths, and effectively 
implements an ensemble average, as we have also 
explictly verified by performing a few statistically independent runs. 

%
\subsection{Exponential growth of gauge-field energy density}
\la{ss:growth}

%
\begin{figure}[t]
    \centering
    \begin{subfigure}[c]{0.52\textwidth}
    
        \hspace*{-0.8em}\includegraphics[width=0.96\textwidth]%
        {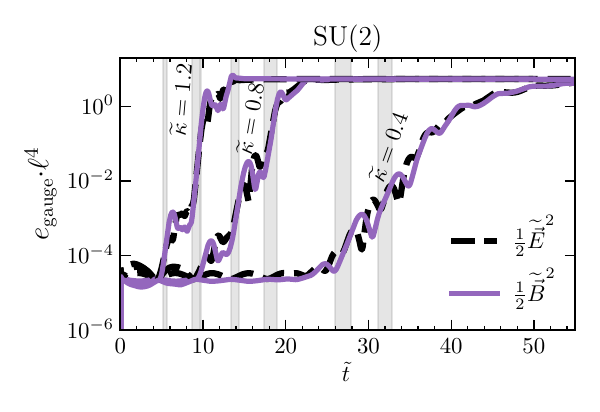}%
        
        \vspace{-0.9em}
        
        \hspace*{-0.8em}\includegraphics[width=0.96\textwidth]%
        {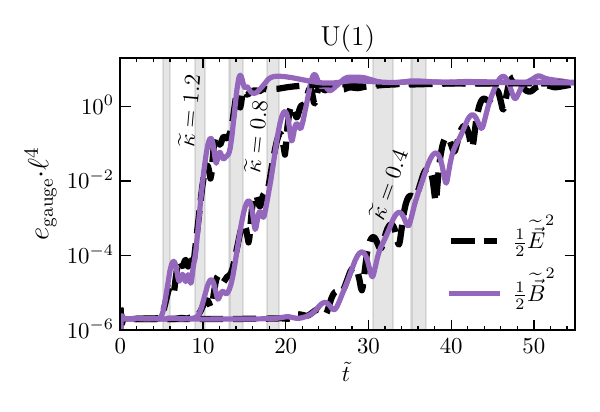}%

    \end{subfigure}%
    \hspace{0.01\textwidth}%
    \begin{subfigure}[c]{0.47\textwidth}

        \vspace{1.3em}
        \hspace{-1.5em}\includegraphics[width=1.06\textwidth]%
        {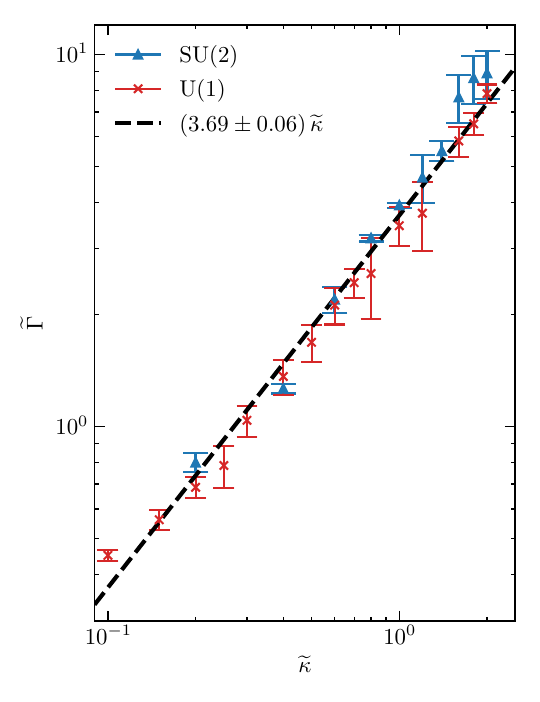}

    \end{subfigure}%

    \vspace*{-4mm}
    
    \caption[a]{\small Left: 
    Evolution of gauge-field energy densities 
    as a function of time, $\tilde{t}$, 
    for several $\widetilde\kappa$, 
    with the vector notation denoting a sum over
    both spatial and colour indices. \rev{In these plots, 
    $\ell \equiv m^{-1}_{ }$.}
    The grey bands 
    indicate domains of exponential growth. 
    \rev{Comparisons with
    the condensate energy density and with that of axion fluctuations
    can be found in \figs\ref{fig:equi} and \ref{fig:zero-mode}.}
    Right: The coefficient characterizing the 
    exponential growth, $\widetilde\Gamma$ (cf.\ \eq\nr{def_Gamma}),
    as a function of $\widetilde{\kappa}$.
    }
    \label{fig:gauge_growth}
\end{figure}
%

Linear perturbation theory, 
as explained around \eq\eqref{B_modes}, indicates that if
$
 \widetilde{\kappa} \dot{\bar{\widetilde{\varphi}}}   \neq 0
$, 
then IR modes with 
$
 \tilde{k} < 
  \widetilde{\kappa} | \dot{\bar{\widetilde{\varphi}}} |^{ }_\rmi{max}
$ can experience exponential growth. 
The results of our numerical simulations 
validate this expectation. In \fig\ref{fig:gauge_growth} we show 
the evolution of the energy densities in the electric 
and magnetic fields as a function of time, for several values 
of $\tilde{\kappa}$. The grey bands indicate regions of exponential 
growth. Between the grey bands, the growth ceases for
a while; this corresponds to $ |\dot{\bar{\widetilde{\varphi}}}| $ 
becoming smaller, so that the instability band becomes narrow. 
This general behaviour is in accordance with \fig\ref{fig:modes}. 
After having gone through 
a few instability bands, the gauge field energy density
\rev{becomes constant}; this takes longer for small $\widetilde{\kappa}$, 
as the instability band is narrower. 
Then also the oscillations of $ \bar{\widetilde{\varphi}} $ 
continue longer (cf.\ \fig\ref{fig:oscillations}). 

In the grey bands of \fig\ref{fig:gauge_growth}, 
in which the relative growth rate
of the gauge field energy density, 
$\dot{e}^{ }_\rmi{gauge}/e^{ }_\rmi{gauge}$,  
is approximately constant, we fit the behaviour to 
\begin{equation}
 e_{\text{gauge}} (\,\tilde{t}\,)
 \; \approx \;   
 e_{\text{gauge,0}} \; 
 e^{\tilde{\Gamma} \Delta \tilde{t}}_{ }
 \;, \quad
 \Delta\tilde t \; \equiv \; \tilde{t} - \tilde{t}^{ }_0  \, .
 \la{def_Gamma}
\end{equation}
To obtain an uncertainty estimate 
for $\tilde{\Gamma}$, we average it 
over several bands. In the right panel 
of \fig\ref{fig:gauge_growth}, we plot $\tilde{\Gamma}$
as a function of $\tilde{\kappa}$ both for SU(2) and U(1). 
We observe an approximately linear dependence (dashed line), 
whose coefficient is similar in both cases. 

%
\subsection{Damping of axion oscillations}
\la{ss:damping}

%
\begin{figure}[t]
    \centering
    \begin{subfigure}{0.49\textwidth}
        \hspace{-2.5em}\includegraphics[width=1.1\textwidth]%
         {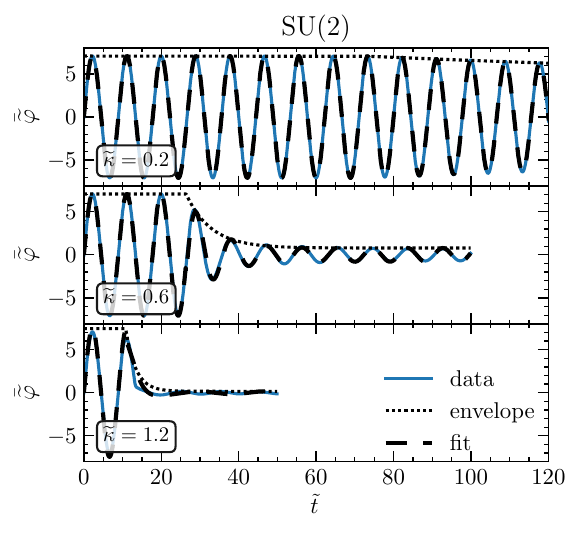}%
    \end{subfigure}%
    \begin{subfigure}{0.49\textwidth}
        \hspace{-1em}\includegraphics[width=1.1\textwidth]%
         {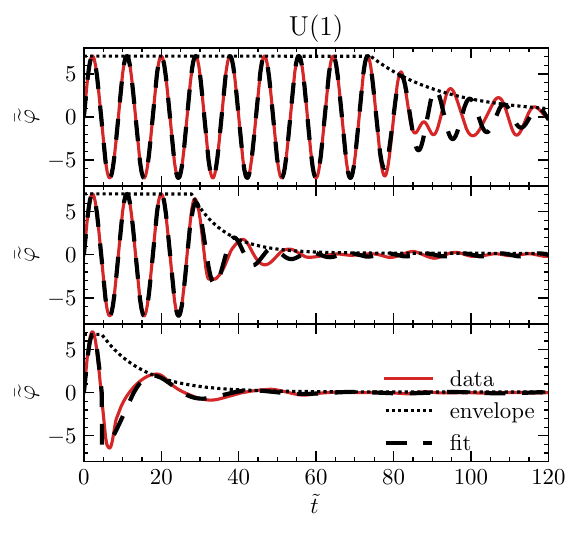}%
    \end{subfigure}%
    \caption[a]{\small Illustration of axion oscillations
    and their subsequent damping, for SU(2) (left) and U(1) (right), 
    for three representative values of $\widetilde\kappa$. 
    The dotted lines show a fit to the envelope of \eq\nr{ansatz}, 
    whereas the dashed lines represent a fit to the full \eq\nr{ansatz}.
    }
    \label{fig:oscillations}
\end{figure}
%

Turning to the axion field, from the measured data we extract
the axion condensate through a spatial average, 
\be
  \bar{\!\widetilde\varphi}(\,\tilde t\,)
  \; \equiv \; 
  \frac{1}{N^3_{ }}
  \sum_{\vec{x}}
  \widetilde\varphi (\,\tilde t\,,\vec{x})
  \;. \la{def_bar_varphi}
\ee
Its evolution is shown in \fig\ref{fig:oscillations}, revealing
a long period of oscillations for $\widetilde \kappa \ll 1.0$, 
and an overdamped regime for $\widetilde \kappa \gg 1.0$. Recalling
\fig\ref{fig:gauge_growth}, the damping 
is due to a transfer of energy from the
condensate to the gauge degrees of freedom
(and to condensate fragmentation). 

On a closer inspection, the pattern of damping is 
non-trivial, and only sets in after a while. 
In the SU(2) case we find that, as illustrated by the dashed
lines in \fig\ref{fig:oscillations}  (left), the data can 
be well represented by \eq\nr{ansatz}.
Here, $\mathcal{A}_0$ and $\mathcal{A}$ 
describe the initial and final oscillation amplitudes, respectively. 
The parameter $\tilde t^{ }_0$ describes the period of unattenuated 
oscillations, $\widetilde\omega_{1}^{ }$  and $\widetilde\omega_{2}^{ }$
the oscillation frequencies before and after the start of damping,
and $\widetilde\Upsilon$ the damping coefficient. 
The phase factor~$\phi_2^{ }$ is matched
to guarantee continuity
(the first derivative has a jump).
The original oscillation frequency, $\widetilde\omega_1^{ }$, 
is close but not equivalent to its tree-level value, $\widetilde m$. 
We find that after the damping starts, 
the oscillations slow down, 
$\widetilde\omega_2^{ } < \widetilde\omega_1^{ }$, 
\rev{
in accordance with the behaviour expected from
a damped harmonic oscillator.}

We interpret the delay parameter, 
$\tilde t^{ }_0$, as originating from that the gauge modes 
need to grow to a critical threshold before allowing 
for a significant backreaction. As shown in 
\fig\ref{fig:oscillations}  (right), the delay is also present in U(1), 
but the subsequent damping is not well described by \eq\nr{ansatz}. 
Hence, the physics leading to \eq\nr{ansatz} must reflect non-Abelian
gauge self-interactions, for instance processes
resembling sphaleron damping.


%
\begin{figure}[t]
    \centering
    \includegraphics[width=0.55\textwidth]{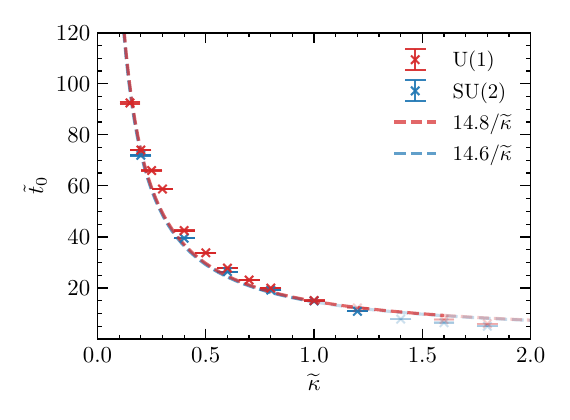}%
    \caption[a]{\small
      Illustration of the coefficient $\tilde t^{ }_0$,
      characterizing the delay of when damping starts,
      as extracted from a fit to \eq\nr{ansatz}.
      With a lighter shade we show points in the overdamped regime,
      where less than a full oscillation occurs before the damping sets in.
      In the latter regime, the error of the fitting procedure is significant.
      }
    \label{fig:t_0}
\end{figure}
%

Let us take a closer look at $\tilde t^{ }_0$.
As shown in \fig\ref{fig:t_0}, 
its dependence on $\widetilde\kappa$ can be well represented 
by $\tilde t^{ }_0 \propto 1/\widetilde\kappa$.
The value of $\tilde{t}_0$ 
asymptotes to zero for large $\widetilde{\kappa}$, though it must
be added that as $\tilde t^{ }_0\to 0$, 
less than a single undamped oscillation is present, 
which significantly increases the systematic uncertainty 
of our fitting routine. We indicate this by the lighter shading 
of these data points in our plot.


%
\begin{figure}[t]
    \centering
    \includegraphics[width=0.55\textwidth]{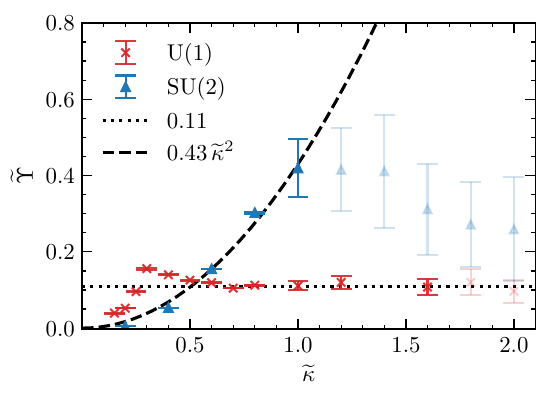}%
    \caption[a]{\small 
   The axion damping 
   coefficient $\widetilde\Upsilon$, as extracted from
   a fit to \eq\nr{ansatz}, for both SU(2) and U(1).
   We also illustrate possible quadratic and constant representations
   of the data. 
   The points in the overdamped regime are shown with a lighter shade.
   Here less than a full oscillation takes place before
   the damping sets in, which significantly adds to the error
   of the fitting procedure.
   }
    \label{fig:Upsilon}
\end{figure}
%

Turning to the axion damping coefficient, $\widetilde\Upsilon$, 
shown in \fig\ref{fig:Upsilon}, we find that a quadratic behaviour 
fits the SU(2) results for modest $\tilde\kappa$. This suggests that 
gauge fields are produced from the axion condensate 
at $\rmO(\widetilde\kappa)$, and backreact on it 
at $\rmO(\widetilde\kappa^2_{ })$. This is in analogy with
sphaleron damping, which is also of $\rmO(\widetilde\kappa^2_{ })$.

However, 
when $\widetilde\kappa$ is large,
we enter the overdamped regime, 
where less than a single oscillation is present. 
In this case there is an increased uncertainty associated 
with fitting $\tilde{t}_0$ and $\widetilde{\Upsilon}$. 
To approximate the systematic error,
we determine $\widetilde{\Upsilon}$ for different fixed 
$\tilde t_0^{\hspace*{0.3mm}\prime}
\in (\tilde t_0^{ } - \pi/\tilde m,\tilde t_0^{ } + \pi/\tilde m)$.
As visible in \fig\ref{fig:Upsilon}
(points with lighter shade), the uncertainty soon becomes of order unity.

In the U(1) case, 
we observe first an increase of~$\widetilde\Upsilon$ with~$\widetilde\kappa$, 
and then a quick flattening to a constant value.
In a different context, a similar flattening 
is observed in \fig\ref{fig:equi}.
It is conceivable that the U(1) dynamics is dictated 
by the tachyonic instability, whereas in the SU(2) case, 
gauge-field self-interactions moderate the instability and 
thereby regulate the dynamics, leading to a ``sequential'' behaviour, 
associated with identifiable powers of $\widetilde\kappa$. 

%
\subsection{Power spectra and energy equipartition}
\la{ss:equi}

%
\begin{figure}[t!]
\centering

 \vspace*{-4mm}

   \begin{subfigure}[b]{1.0\textwidth}
 \includegraphics[width=1\textwidth]{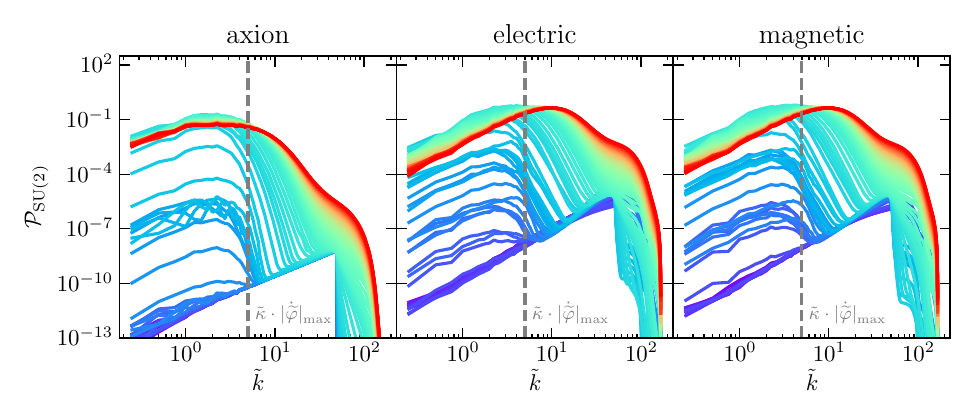}
   \label{fig:Ng1} 
 \end{subfigure}

 \vspace*{-7mm}

 \begin{subfigure}[b]{1.0 \textwidth}
 \includegraphics[width=1\textwidth]{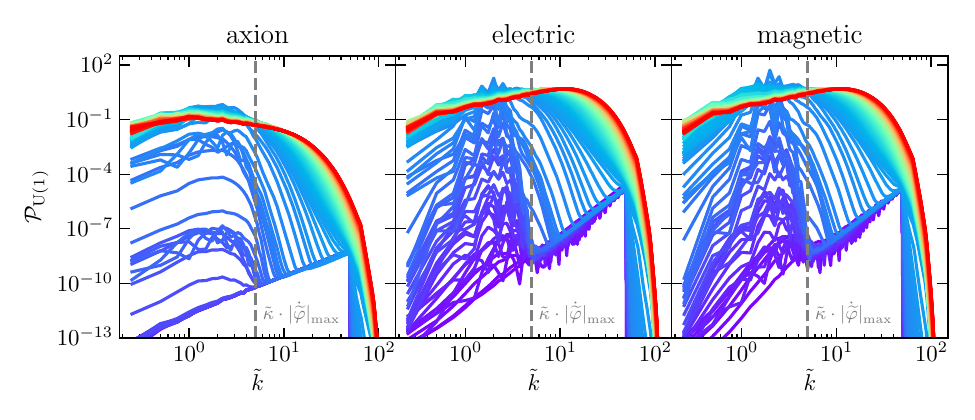}
   \label{fig:Ng2}
 \end{subfigure}

 \vspace*{-7mm}

\caption[a]{\small
 Top panels: 
 Time evolutions of the SU(2) power spectra for 
      ${\widetilde\varphi}$, 
      ${\widetilde E}$, and  
      ${\widetilde B}$,  
      for $\widetilde\kappa = 1.0$, 
      at time intervals $\Delta\tilde t = 0.5$.
 Bottom panels: 
 The same for U(1). 
 On the qualitative level, 
 the right panels can be compared with the linearized
 expectations from \fig\ref{fig:modes}.
}
\label{fig:power_spectra}
\end{figure}
%

In analogy with \eqs\nr{P_cl} and \eqref{B_P}, 
we define electric and the axion power spectra as 
\be
 \P^{ }_{\tilde E}
 \; 
 {\equiv}
 \; 
 \frac{ \dA^{ } \tilde k^3_{ } }{2 \pi^2_{ }}
 \bigl[\, 
   |\widetilde E^{ }_{k,+}(\tilde t \hspace*{0.3mm})|^2_{ }
  + 
   |\widetilde E^{ }_{k,-}(\tilde t \hspace*{0.3mm})|^2_{ }
 \,\bigr]
 \;, \quad
 \P^{ }_{\tilde \varphi}
 \; 
 {\equiv}
 \; 
 \frac{  \tilde k^3_{ } }{2 \pi^2_{ }}
   |\widetilde \varphi^{ }_k (\tilde t \hspace*{0.3mm})|^2_{ }
  \la{P_varphi} \, , 
\ee
where $\dA^{ } = 3\, (1)$ 
denotes the number of generators of the gauge algebra,
and $\widetilde E^{ }_{k,\pm}$ and $\widetilde\varphi^{ }_k$
are mode functions (if one uses Fourier transforms instead, 
the normalization contains an additional factor, 
cf.\  \eq\nr{fourier_norm}). 
In \fig\ref{fig:power_spectra} we show their time evolution
for $\widetilde\kappa = 1.0$, from an initial time 
$\tilde{t} = 0$ to a final 
time $\tilde{t} = 50$. Dark blue curves correspond to early times, 
dark red curves to late times. 
The sharp cutoff of the initial spectrum is due to the imposed 
momentum cutoff $\Lambda$ in \eq\nr{variance_new}. 

%
\begin{figure}[t]
    \centering

    \includegraphics[width=0.95\textwidth]%
    {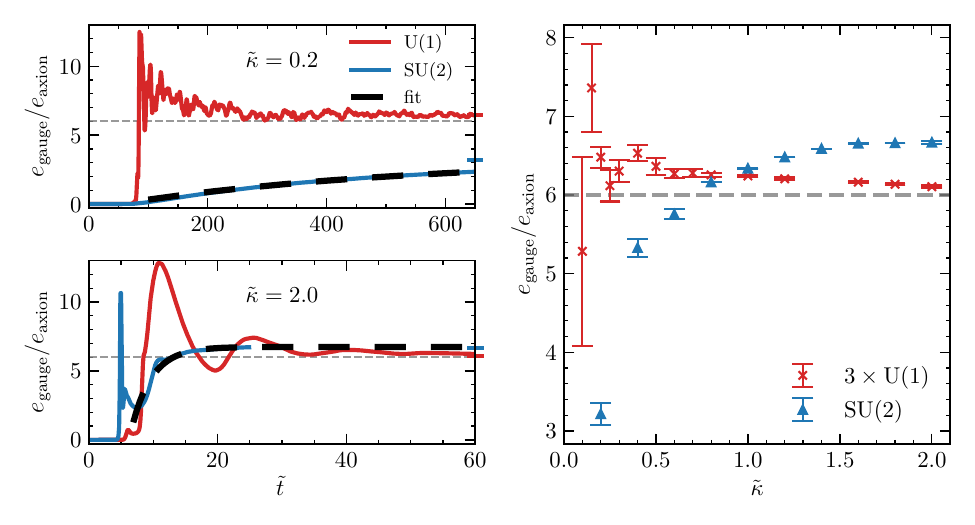}%

    \caption[a]{\small
     Left: The time evolution of the ratio of the gauge and axion 
     energy densities, for  small and large $\widetilde\kappa$.
     The U(1) results have been multiplied by a factor~$3$. For small 
     $\widetilde\kappa$, the energy transfer to the gauge sector is 
     much faster for U(1), due to the tachyonic instability. 
     For SU(2), the dashed lines shown an exponential fit to late times. 
     The asymptotic late-time values are indicated with
     short blocks along the right vertical axis. 
     Right: The 
     late-time values as a function of $\widetilde\kappa$.
     }

    \label{fig:equi}
\end{figure}
%

%
\begin{figure}[t]
    \centering
    \includegraphics[width=0.99\textwidth]{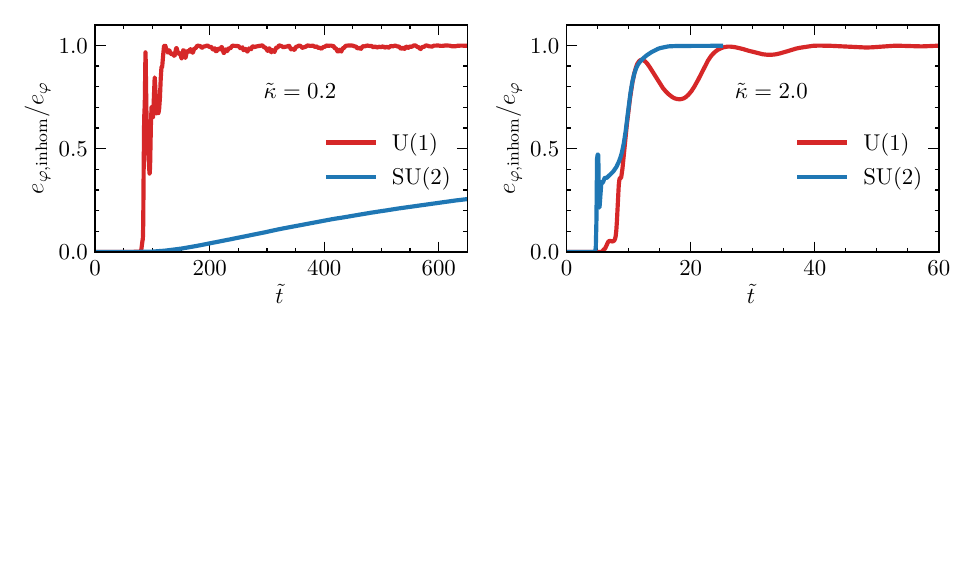}%
    \vspace*{-40mm}%
    \caption[a]{\small
    The fraction of axion energy density in inhomogeneous fluctuations,
    for $\widetilde\kappa = 0.2$ (left) and $\widetilde\kappa = 2.0$ (right). 
    Comparing with \fig\ref{fig:equi}, the transfer of energy to axion 
    inhomogeneities happens approximately simultaneously with 
    the transfer of energy to the gauge sector.
 	}

    \label{fig:zero-mode}
\end{figure}
%

The power spectra show initial growth in the instability band 
$\tilde k < \widetilde{\kappa} |\dot{\bar{\widetilde{\varphi}}}|^{ }_{\text{max}}$, 
indicated with the dashed vertical lines. 
Subsequently energy gets 
transferred to higher-momentum modes.
The process is noticeably smoother in SU(2) 
(top panels) than in U(1) (bottom panels), 
with the latter
displaying oscillatory features reminiscent of \fig\ref{fig:modes}.
We interpret the smoothening of the SU(2) power spectra as originating
from the gauge self-interactions present in that case. However, 
there are interactions also in the U(1) case, mediated by the axions. 
Therefore, in both cases, energy is ultimately 
transferred also outside
of the instability band. In fact, in the end, 
much of the power lies in the
UV modes, which in the classical lattice theory reflect dispersion
relations of the Brillouin zone. 
 
In the left panels of \fig\ref{fig:equi}, 
we quantify energy equipartition between the gauge and 
axion sectors, as a function of time, for two values
of the coupling strength 
$\widetilde{\kappa}$. For SU(2), where the energy transfer is 
fairly slow for small $\widetilde\kappa$, the estimate
for the asymptotic value 
is obtained from an exponential fit at finite times. 
For U(1), the process is much faster, and the asymptotic value
is estimated by taking an average over oscillations at late times. 

The asymptotic energy densities are shown 
in the right panel of \fig\ref{fig:equi}, where we have summed over 
the electric and magnetic contributions. The U(1) results have been 
multiplied by a factor 3, to make them easily comparable with the SU(2) results. 

For SU(2), we observe a smooth dependence on $\widetilde\kappa$, with
the $\widetilde\kappa\to 0$ limit decreasing, and the 
large-$\widetilde\kappa$ limit overshooting the expected value of 6 by $\sim 10\%$. 
The value~6 corresponds to a non-interacting continuum estimate, 
accounting for the two degrees of freedom per each of the three 
gauge bosons, divided by the one degree of freedom of the axion. 
The deviation from 6 originates likely from interactions,
which modify both the axion and the gauge-field energy densities. 

In contrast, 
in the U(1) case, the dependence on $\widetilde\kappa$ is {\em not} 
smooth, but the energy is efficiently 
transferred to the gauge sector 
as soon as $\widetilde\kappa \neq 0$. The efficient
transfer likely reflects the tachyonic instability. The ratio 
equals~2, accounting for two degrees of freedom 
of the photon divided by  
one degree of freedom of the axion, 
up to a correction of $\sim 3\%$. The 
relative deviation from the free value is smaller
than for SU(2), likely due to the absence of self-interactions
in the gauge sector. 

Finally, in \fig\ref{fig:zero-mode}, we show the fraction of the axion energy density that is contained in inhomogeneous fluctuations, after subtracting the contribution from the condensate, as defined in \eq\nr{def_bar_varphi}, such that 
\ba
	\widetilde{\varphi}_{\text{inhom}}(\tilde{t},\vec{x}) 
    & \equiv & \widetilde{\varphi}(\tilde{t},\vec{x}) 
    \;-\;
    \,\bar{\!\widetilde{\varphi}}(\tilde{t})
   \;, 
   \\[2mm]
	\widetilde e_{\varphi, \text{inhom}}^{ }(\tilde t,\vec{x}) 
    & \equiv & 
     \frac{1}{2}   
	\dot{\widetilde{\varphi}}^{\hspace*{0.3mm}2}_{\text{inhom}}
      (\tilde{t},\vec{x})
    +\frac{1}{2} 
     \left|\,\nabla   \widetilde{\varphi}_{\text{inhom}}
             (\tilde{t},\vec{x}) \,\right|^2_{ }
     + 
     \frac{1}{2} \widetilde m^2_{ } 
     \widetilde{\varphi}^{\hspace*{0.3mm}2}_{\text{inhom}}(\tilde{t},\vec{x})
    \;.
\ea
\rev{The
condensate energy has the same form, but without the spatial 
gradients.}
If we interpret the transfer of energy to the gauge sector 
as a prerequisite for the latter to equilibrate, then this plot shows 
that the axion field has a chance to equilibrate about as fast as 
the gauge plasma. However, we remark that from the point of view of 
warm inflation, the equilibration of the gauge plasma is 
the more important phenomenon, as the gauge plasma dominates 
the radiation energy density and generates the sphaleron friction.

%
\section{Conclusions and outlook}
\la{se:concl}

The purpose of this paper has been to initiate the numerical study
of how an axion-like inflaton field ($\varphi$)
and a non-Abelian gauge ensemble
interact with each other. To this aim, we have varied 
the dimensionless coupling characterizing
the strength of the interaction, 
$
 \widetilde\kappa = \alpha\hspace*{0.3mm} m / (2 \pi f^{ }_a )
$
(cf.\ \eqs\nr{kappa} and \nr{rescale0}), 
in the range 
$
 \widetilde\kappa = 0.2, ... , 2.0
$, 
and simulated the system on a lattice of 
size $756^3_{ }$. Most of the initial energy
density is carried by a spatially homogeneous $\varphi$-condensate.
Adding a tiny amount of noise switches on the decay channel
to the gauge ensemble, and the energy density of
the condensate
gets rapidly transferred to other degrees 
of freedom (gauge fluctuations 
and spatial variations of $\varphi$). 

The dynamics that we observe for SU(2) displays smooth
parametric dependences, characteristics
of a linear-response regime. 
The energy density in the 
electric and magnetic fields grows exponentially in time, 
with a growth rate approximately linearly proportional 
to $\widetilde\kappa$,
\be
 \widetilde\Gamma 
 \; 
 \overset{\rmii{\fig\ref{fig:gauge_growth}\vphantom{\Big | }}}{\approx} 
 \; 
 3.7\, \widetilde\kappa 
 \quad
 \overset{\rmii{\nr{rescale0}\vphantom{\Big | }}}{\Rightarrow} 
 \quad
 \Gamma 
 \; \approx \; 
 3.7 \, m^2_{ }\, \kappa 
 \;. \la{Gamma}
\ee
Axion oscillations get damped, and in the SU(2) case,
the damping rate is approximately quadratic in $\widetilde\kappa$, 
\be
 \widetilde\Upsilon 
 \overset{\rmii{\fig\ref{fig:Upsilon}\vphantom{\Big | }}}{\approx} 
 0.4 \, \widetilde\kappa^{\hspace*{0.3mm}2}_{ } 
 \quad
 \overset{\rmii{\nr{rescale0}\vphantom{\Big | }}}{\Rightarrow} 
 \quad
 \Upsilon 
 \; \approx \; 
 0.4 \, m^3_{ }\, \kappa^2_{ } 
 \;. \la{Upsilon}
\ee 
However, axion damping only sets in after a while, 
once the gauge ensemble has obtained 
a sufficient energy density, and this delay factor
scales as
\be
 \tilde t^{\hspace*{0.3mm}-1}_0 
 \overset{\rmii{\fig\ref{fig:t_0}\vphantom{\Big | }}}{\approx} 
 0.07\, \widetilde \kappa^{ }_{ }
 \quad
 \overset{\rmii{\nr{rescale0}\vphantom{\Big | }}}{\Rightarrow} 
 \quad
 t^{\hspace*{0.3mm}-1}_0 
 \; \approx \; 
 0.07\, m^2_{ }\, \kappa^{ }_{ }  
 \;. \la{t_0} 
\ee 
The delay may be sensitive to the amount of noise
that we inserted in the initial state, and therefore not 
an equally important physical characteristic of the system
as $\Gamma$ and $\Upsilon$.
In the end, the overall energy density in the gauge sector 
is about $2(\Nc^2 - 1)$ times larger than that in the axion field, 
manifesting energy equipartition (cf.\ \fig\ref{fig:equi}). 

Physically, it would be tempting to identify the energy
equipartition that we observe
with equilibration (or thermalization). 
However, strictly speaking
the latter phenomenon cannot be studied with classical
lattice gauge theory, 
because the thermal momentum scale, $k\sim T$, has been 
replaced by the unphysical lattice cutoff scale, $k \sim \pi/a$. 
Nevertheless, given that overall energy 
is conserved and the Hubble rate vanishes, 
the physical analogue of our setup corresponds to 
{\em instantaneous reheating}, in which 
all of the condensate energy density gets ultimately 
transferred to equipartitioned UV degrees of freedom. 

Turning to a comparison with 
the U(1) case, we note that even though it does 
display many similarities with SU(2), 
we have also shown that there is a very marked difference.
In U(1), the dependence on $\kappa$ is not smooth, but even
a small non-zero value leads to rapid equipartition
(cf.\ \fig\ref{fig:equi}). We suspect that this is due to 
the tachyonic instability, which is not much attenuated 
in the absence of gauge-field self-interactions. 

For the present study, 
we have kept the non-Abelian
gauge coupling fixed, $g^2_{ } = 1.0$. 
In principle, it would be interesting to vary the strength
of the gauge self-interaction, while keeping $\kappa$ fixed. 
Unfortunately
it is not entirely trivial to do this in the classical theory, 
due its scale invariance.\footnote{%
 The evolution equations, 
 \nr{eom_latt_scalar_resc}--\nr{eom_Ei_resc}, are invariant
 under the simultaneous change
 $g\to x\, g$, 
 $\widetilde\kappa \to x\, \widetilde\kappa$, 
 $\widetilde\varphi\to \widetilde\varphi/x$, 
 $\widetilde E^a_i\to \widetilde E^a_i/x$, 
 and $\widetilde B^a_i\to \widetilde B^a_i/x$.
 Therefore any change of $g$ can be compensated for
 by a change of $\widetilde\kappa$ and
 a rescaling of initial conditions.   
}
Nevertheless, it would be interesting to explore how this
variation could be implemented in a physically meaningful way. 

A second important next step is to include the expansion of the universe
in the simulations. The expansion option is built in into the 
\href{https://github.com/cosmolattice/cosmolattice}{CosmoLattice}
platform~\cite{cosmo1,cosmo2}, 
so we hope to return to this topic in the foreseeable future. 

Finally, the inclusion of fermions represents another interesting avenue.
Even if elementary fermions do not permit for a classical description, an ensemble average
over fermions
can indeed be included, via a time-dependent chemical potential, 
which can bias 
sphaleron processes. We leave this extension to an upcoming study. 

%
\section*{Acknowledgements}
KB thanks the U.S. Department of Energy, Office of Science, 
Office of High Energy Physics, under Award Number DE-SC0011632, 
and the Walter Burke Institute for Theoretical Physics.
Test runs for this work were conducted at the Resnick High Performance 
Computing Center, a facility supported by Resnick Sustainability Institute 
at the California Institute of Technology. AF and FRS are supported 
by the Deutsche Forschungsgemeinschaft (DFG, German Research Foundation) 
through the Emmy Noether Programme Project No.\ 545261797. 
The authors gratefully acknowledge 
the computing time made available to them on the high-performance 
computer Noctua 2 at the NHR  Paderborn Center for Parallel Computing (PC2). 
This center is jointly supported by the Federal Ministry of Research, 
Technology and Space and the state governments participating 
in the National High-Performance Computing (NHR) joint 
funding program (www.nhr-verein.de/en/our-partners).

%
\appendix
\renewcommand{\thesection}{\Alph{section}} 
\renewcommand{\thesubsection}{\Alph{section}.\arabic{subsection}}
\renewcommand{\theequation}{\Alph{section}.\arabic{equation}}

\newpage

%
\section{Equations of motion in continuum}
\la{se:cont}

Though well known, we specify here
the continuum evolution equations and initial conditions that our
study corresponds to. For ease of reference, 
the discussion is structured identically to \app\ref{se:latt}, 
where the technically more involved discretized setup is introduced.  

%
\paragraph{Notation.}

It is convenient to define a covariant derivative 
as 
$
 D^{ }_\mu \equiv \partial^{ }_\mu + i g T^a_{ } A^a_\mu{ }
$,
given that this removes unnecessary minus signs from the lattice
formulation (cf.,\ e.g., \eqs\nr{link}--\nr{time_derivative}).  
Here $T^a_{ }$ are the generators of SU($\Nc^{ }$), 
normalized as
$
 \tr(T^a_{ }T^b_{ }) = \delta^{ab}_{ }/2
$.
Then the components of the field strength, 
$
 F^{ }_{\mu\nu} \equiv T^a_{ } F^a_{\mu\nu} 
 = [D^{ }_\mu,D^{ }_\nu]/(ig)
$,
can be written as
\be
 F^{a}_{\mu\nu} = \partial^{ }_\mu A^a_\nu - 
 \mathcal{D}^{ac}_\nu A^c_\mu
 \;, \la{Fmunu}
\ee
where 
$
  \mathcal{D}^{ac}_\nu \equiv
  \delta^{ac}_{ }\partial^{ }_\nu - g f^{abc}_{ }A^b_\nu
$
is the covariant derivative in the adjoint representation. 

We express the components of the field strength as 
\be
 F^a_{0i}
 \; \overset{\rmii{\nr{Fmunu}}}{=} \;
 \underbrace{
 \dot{A}^a_{i}
 }_{\,\equiv\, E^a_i 
   }
 - \mathcal{D}^{ac}_{i}A^c_0
 \;, \quad
 F^a_{ij}
 \; \equiv \;
 \epsilon^{ }_{ijk} B^a_k
 \;. \la{F0i_full}
\ee
This defines the electric and magnetic fields that we 
make use of in the following. 
We note that when $\kappa\neq 0$ (cf.\ \eq\nr{kappa}), 
the electric field does {\em not}
correspond to a canonical momentum.

Making use of $\epsilon^{0ijk}_{ } = \epsilon^{ }_{ijk}$, 
the Lagrangian, $L = \int_\vec{x}\mathcal{L}$, where 
$\mathcal{L}$ is from \eq\nr{L}, 
can be written as 
\be
 {L} 
 \;
 \underset{\rmii{\nr{kappa}}}{
 \overset{\rmii{\nr{L}}}{=}}
 \;
 \int_\vec{x} 
 \biggl\{ 
     \frac{1}{2} \dot{\varphi}^2_{ }
    - \frac{1}{2}  \partial^{ }_i\varphi\, \partial^{ }_i \varphi
    - V(\varphi)
    + \frac{1}{2} F^a_{0i}F^a_{0i}
    - \frac{1}{4} F^a_{ij}F^a_{ij}
    - \kappa\, \varphi\, F^a_{0i} B^a_i
 \biggr\}
 \;. \la{L_expl} 
\ee

%
\paragraph{Gauss constraint and Hamiltonian.}

Given that the Lagrangian in \eq\nr{L_expl}
does not depend on $\dot{A}^a_0$, 
the corresponding Euler-Lagrange equation states that 
\be
 0 \;=\;
  \frac{{\rm d}}{{\rm d}t} \frac{\delta L}{\delta \dot{A}^a_0(x)}
  \; =\; 
 \frac{\delta L}{\delta {A}^a_0(x)}
 \;, \quad
 x \;\equiv\; (t,\vec{x})
 \;. 
\ee
Subsequently, we choose the temporal gauge, 
$
 A^a_0(x) = 0
 \quad
 \forall a,x
$.
The field $A^a_0$
appears in \eq\nr{L_expl} only through the field strength $F^a_{0i}$, 
cf.\ \eq\nr{F0i_full}. Thus, 
\ba
 0 & = & G^{ }_{ }(x) 
 \; \equiv \;  
 T^a_{ }\frac{\delta L}{\delta A^a_0(x)} \bigg|^{ }_{A^a_0 = 0}
 \; = \; 
 T^a_{ }
 \,\bigl[\, 
 \mathcal{D}^{ac}_i E^c_i -\kappa\, \mathcal{D}^{ac}_i (\varphi B^c_i)
 \,\bigr]
 \;. \la{Gauss_cont_pre}
\ea
Making use of the Jacobi identity
\be
 T^a_{ } \mathcal{D}^{ac}_i B^c_i 
 \;=\;
 \epsilon^{ }_{ikm} [D^{ }_i,[D^{ }_k,D^{ }_m]]/(2ig)
 \;=\; 0 
 \;, 
 \la{jacobiI_pre}
\ee
which implies that $\mathcal{D}^{ac}_i$ only acts on $\varphi$
in \eq\nr{Gauss_cont_pre}, 
leads to the Gauss constraint, which is given in \eq\nr{Gauss_cont}.
The Jacobi identity itself 
is expressed more concisely through \eq\nr{jacobiI}.

Apart from the Lagrangian, 
we are interested in the total energy of the system. 
We refer to this as the {\em Hamiltonian}, $H$, 
even though
we do {\em not} re-express the time derivatives (``$\dot{q}$'')
in terms of canonical momenta (``$p$''). 

The Hamiltonian is obtained with a Legendre transform from the 
Lagrangian: 
\be
 H \; = \; 
 \int_\vec{x}
 \biggl[\,
 \dot{\varphi}(x) 
 \frac{\delta L  |^{ }_{A^a_0 = 0} }{\delta \dot\varphi(x)}
 + 
 E^a_i(x)
 \frac{\delta L  |^{ }_{A^a_0 = 0} }{\delta E^a_i(x)}
 \,\biggr]
 \; - \; L \big|^{ }_{A^a_0 = 0}
 \;. \la{Legendre}
\ee
The axion-gauge coupling drops out in this transformation, 
and we obtain
\be
 {H}
 \; 
 \underset{\rmii{\nr{Legendre}}}{
 \overset{\rmii{\nr{L_expl}}}{=}}
 \;
 \int_\vec{x} 
 \biggl\{ 
     \frac{1}{2} \dot{\varphi}^2_{ }
    + \frac{1}{2}  \partial^{ }_i\varphi\, \partial^{ }_i \varphi
    + V(\varphi)
    + \frac{1}{2} E^a_{i}E^a_{i}
    + \frac{1}{4} F^a_{ij}F^a_{ij}
 \biggr\}
 \;. \la{H_expl}
\ee

%
\paragraph{Evolution equations.}

After having fixed $A^a_0 = 0$ and taken care of the consistency
of this procedure through the Gauss constraint, 
the remaining Euler-Lagrange equations read
\be 
 \partial^{ }_t \biggl( \frac{\delta L  |^{ }_{A^a_0 = 0} }
 {\delta\dot\varphi(x)} \biggr)
 \;=\; 
 \frac{\delta L  |^{ }_{A^a_0 = 0} }{\delta\varphi(x)}
 \;, \quad
 \partial^{ }_t 
 \biggl( 
 \frac{\delta L  |^{ }_{A^a_0 = 0} }
 {\delta E^a_i(x)} 
 \biggr)
 \;=\; 
 \frac{\delta L |^{ }_{A^a_0 = 0} }{\delta A^a_i(x)}
 \quad
 \forall i,a,x
 \;. \la{E-L}
\ee
{}From \eq\nr{L_expl}, 
the derivative with respect to $E^a_i$ yields
\be
 \frac{\delta L  |^{ }_{A^a_0 = 0} }
 {\delta E^a_i(x)} 
 \; 
 \overset{\rmii{\nr{L_expl}}}{=} 
 \; 
 E^a_i - \kappa \, \varphi\, B^a_i  
 \;. \la{der_Eai}
\ee
For the derivative with respect to $A^a_i$, we note that
\be
 \frac{\delta}{\delta A^a_i(x)} 
 \int_\vec{y} F^{b}_{jk}\, f(y) 
 \; = \; 
 \bigl( \delta^{ }_{i,j} \mathcal{D}^{ab}_k - 
        \delta^{ }_{i,k} \mathcal{D}^{ab}_j \bigr) f(x)
 \;. \la{der_A}
\ee
This gives
\be
 \frac{\delta L |^{ }_{A^a_0 = 0} }{\delta A^a_i(x)}
 \;
 \underset{\rmii{\nr{der_A}}}{
 \overset{\rmii{\nr{L_expl}}}{=}}
 \; 
 -\mathcal{D}^{a c}_k F^c_{i k} 
 - \kappa  \epsilon^{ }_{i k j} \mathcal{D}^{a c}_k (\varphi E^c_j )
 \; = \; 
 - \bigl[\,
  \mathcal{D}\times \vec{B}
  + \kappa\, 
  \mathcal{D}\times (\varphi \vec{E})
 \,\bigr]^a_i
 \;. \la{der_Aai}
\ee
When combining \eqs\nr{der_Eai} and \nr{der_Aai} according to 
\eq\nr{E-L}, we can furthermore make use of the Jacobi identity
\be
 \epsilon^{n\mu\nu\rho}_{ } 
 [D^{ }_\mu,[D^{ }_\nu,D^{ }_\rho]]  = 0  \quad \forall n \in \{1,2,3\}
 \;. \la{jacobiII_pre}
\ee
Noting that one among the indices $\mu,\nu,\rho$ is necessarily time, 
and rewriting the identity with the $\vec B$ and $\vec E$ fields, 
this implies the second relation given in \eq\nr{jacobiI}. 

With the help of \eq\nr{jacobiI}, 
we note that $\dot{\vec{B}}$ from 
the time derivative of \eq\nr{der_Eai} and 
$\mathcal{D}\times\vec{E}$ from \eq\nr{der_Aai} 
cancel each other.
To summarize, the equations of motion
obtained from the $A^a_0 = 0$ gauging of \eq\nr{L_expl} 
can be expressed as 
\ba
 \quad
 \ddot{\varphi} 
 & 
 \underset{\rmii{\nr{E-L}}}{
 \overset{\rmii{\nr{L_expl}}}{=}} 
 &
 \nabla^2_{ }\varphi - V'(\varphi) - \kappa \,\vec{E}\cdot\vec{B}  
 \;, \la{eom_cont_scalar} \\[2mm]
 \dot{\vec{A}} 
 & 
 \overset{\rmii{\nr{F0i_full}}}{=} 
 & 
 \vec{E} 
 \;, \la{eom_cont_Ai} \\[2mm]
 \dot{\vec{E}}
  &
  \underset{\rmii{\nr{jacobiI},\nr{E-L}}}{
  \overset{\rmii{\nr{der_Eai},\nr{der_Aai}}}{=}}
  & 
 - 
  \mathcal{D}\times \vec{B}
 + \kappa 
 \bigl(\, 
   \dot\varphi\, \vec{B}
  - \nabla\varphi \times \vec{E}
 \,\bigr)
 \;. \quad 
 \la{eom_cont_Ei} 
\ea

%
\paragraph{Energy conservation.}

A useful crosscheck of the equations of motion is that the total
energy from \eq\nr{H_expl} remains conserved. Expressing $F^a_{ij}$
in terms of $B^a_k$, we obtain
\ba
 \dot{H} 
 &
 \overset{\rmii{\nr{H_expl}}}{=} 
 &  
 \int_\vec{x}
 \bigl\{\, 
 \dot{\varphi}
 \,\bigl[\, 
   \underbrace{
    \ddot{\varphi}
   }_{\rmii{\nr{eom_cont_scalar}}}
  + V'(\varphi) 
 \,\bigr]
    + 
  \partial^{ }_i\varphi\,\partial^{ }_i\dot{\varphi}
  + E^a_i
    \underbrace{
    \dot{E}^a_i
    }_{\rmii{\nr{eom_cont_Ei}}}
  + B^a_i 
   \underbrace{
   \dot{B}^a_i
   }_{\rmii{\nr{jacobiI}}}
 \,\bigr\}
 \nn[2mm]
 & = & 
 \int_\vec{x}
 \bigl\{\, 
   \partial^{ }_i (\dot{\varphi}\, \partial^{ }_i\varphi)
  \;
  \underbrace{
   - \, \vec{E}\cdot \mathcal{D}\times \vec{B} 
   + \vec{B} \cdot \mathcal{D}\times \vec{E} 
  }_{ \epsilon^{ }_{ijk} \partial^{ }_i(E^a_j B^a_k)}
  \; -  
   \underbrace{
   \kappa\,
   \vec{E}\cdot \nabla\varphi \times \vec{E}
   }_{\rm 0~by~antisymmetry}
 \,\bigr\} 
 \;. 
\ea
The integrand is a total derivative, so $\dot{H}$ vanishes if we 
assume periodic boundary conditions for gauge-invariant quantities. 

%
\paragraph{Conservation of Gauss constraint.}

A final crosscheck is offered by that the Gauss
constraint from \eq\nr{Gauss_cont} is time-translation invariant
at every position $\vec x$. 
Writing the spatial indices explicitly, we get
\ba
 \dot{G}
 &
  \overset{\rmii{\nr{Gauss_cont}}}{=}
 & 
 ig \underbrace{
 [\dot{A}^{ }_i,E^{ }_i] }_{0}
  + 
 \underbrace{
 [D^{ }_i,\dot{E}^{ }_i] }_{\rm insert~\rmii{\nr{eom_cont_Ei}}}
 - 
 \kappa\, \dot{\varphi}^{ }_{,i} B^{ }_i 
 - 
 \kappa\, \varphi^{ }_{,i} \dot{B}^{ }_i
 \nn[2mm]
 & = & 
 - \underbrace{ 
 \frac{ \epsilon^{ }_{ijk}\epsilon^{ }_{kmn} }{2}
 [D^{ }_i,[D^{ }_j,[D^{ }_m,D^{ }_n]]] }_{  \,=\, 0}
 \nn[2mm]
 &  & \; + \,  
 \kappa\,\bigl\{ 
    \dot{\varphi} \underbrace{ [D^{ }_i,B^{ }_i] 
     }_{\stackrel{\rmii{\nr{jacobiI}}}{=}0}
  - \underbrace{ \epsilon^{ }_{ijk} \varphi^{ }_{,ij}
     }_{\stackrel{i \leftrightarrow j}{=}0}
     E^{ }_k
  + \varphi^{ }_{,i} 
    \underbrace{ 
    \bigl(\, 
    \epsilon^{ }_{ijk}  [D^{ }_j,E^{ }_k]
  -  \dot{B}^{ }_i
   \,\bigr) 
     }_{\stackrel{\rmii{\nr{jacobiI}}}{=}0}     
 \bigr\}
 \; = \; 
 0 
 \;. 
\ea
The conservation of the Gauss constraint 
is a consequence of many non-trivial 
cancellations, and therefore offers for a strong crosscheck.

%
\paragraph{Initial conditions and power spectra.}

Even though our evolution equations are classical, we can try 
to mimick quantum-mechanical effects 
through initial conditions~\cite{cl1}. Let 
us first recall that if $\varphi$ were a quantum field, denoted by
$\hat\varphi$, it could be
expressed in a mode expansion as 
\be
 \hat\varphi(x)
 \; = \;
 \int_\vec{k} 
 \Bigl[\,
  \hat a^{ }_\vec{k} \, \varphi^{ }_k(t) \, e^{i \vec{k}\cdot \vec{x} }_{ }  
 + 
  \hat a^{\dagger}_\vec{k} \, \varphi^{*}_k(t) 
  \, e^{-i \vec{k}\cdot \vec{x} }_{ }  
 \,\Bigr]
 \;, \quad
 \int_\vec{k} 
  \; \equiv \; 
  \int \! \frac{{\rm d}^3_{ }\vec{k}}{(2\pi)^3_{ }} 
 \;, 
 \la{mode_expansion}
\ee
where the canonical commutation relations and the mode function read
\be
 [\, \hat a^{ }_\vec{k}\,,\, \hat a^{\dagger}_\vec{q}
 \,] 
 \; = \; 
 (2\pi)^3_{ }\delta^{(3)}_{ }(\vec{k-q})
 \;, \quad
 \varphi^{ }_k(t) 
 \; = \; 
 \frac{e^{-i \omega^{ }_k t }_{}}{\sqrt{2 \omega^{ }_k}}
 \;, \quad 
  \omega^{ }_k 
  \; \equiv \; 
  \sqrt{k^2_{ } + m^2_{ }}
 \;. \la{canonical} 
\ee
Computing the equal-time 2-point function in a non-interacting
(``Bunch-Davies'')
vacuum yields
\be
 \langle 0 |\,
   \widehat\varphi(t,\vec x) \,  \widehat\varphi(t, \vec y)
 \,| 0 \rangle
 \; 
 \overset{\rmii{\nr{mode_expansion}}}{=} 
 \; 
 \int_\vec{k} e^{i\vec k\cdot(\vec{x-y}) }_{ }
 \, |\varphi^{ }_k(t)|^2_{ }
 \;. 
 \la{2pt_quantum}
\ee
A quantum-mechanical {\em power spectrum} is defined by
multiplying the integrand of \eq\nr{2pt_quantum}
with the phase-space volume element, 
\be
 \P^{ }_{\varphi}(t,k) 
 \; \equiv \; 
 \frac{k^3_{ }}{2 \pi^2_{ } }
 \, |\varphi^{ }_k(t)|^2_{ } 
 \;. 
 \la{def_P}
\ee

In the classical theory, we write down
an expansion like in \eq\nr{mode_expansion}, but replace
the creation and annihilation operators by complex random variables, 
$c^{ }_\vec{k}$ and $c^*_\vec{k}$. 
This leads to the expressions shown in \eqs\nr{initial_varphi} 
and \nr{initial_dvarphi}. Computing the 2-point function, we 
get a contribution from two terms,  
$
 \langle
 c^{ }_\vec{k}
 c^*_\vec{q}
 \rangle
$
and 
$
 \langle
  c^*_\vec{k}
  c^{ }_\vec{q} 
  \rangle
$.
There is no analogue for the latter in the quantum theory, 
given that $\hat a^{ }_\vec{q} | 0 \rangle = 0$.
To compensate for this, and thereby to reproduce classically
the results in \eqs\nr{2pt_quantum} and \nr{def_P}, we should multiply
the classical noise autocorrelator by the factor $f^{ }_k$, as shown 
in \eq\nr{variance}. However, in the present study, we have reduced
the quantum noise to an even lower level, 
according to \eq\nr{variance_new}.

For the gauge sector, 
we envisage a gauge potential similar to that in 
\eq\nr{initial_varphi}, 
\be
 \vec{A}(t,\vec x)
  \; = \; 
 \int_\vec{k} 
 \sum_{\lambda = \pm}
 \Bigl[\,
   c^{ }_\vec{k} 
  \, \vec{e}_\vec{k}^{\lambda} 
  \, \chi^{ }_k(t) 
  \, e^{i \vec{k}\cdot \vec{x} }_{ }
  +
   c^*_\vec{k} 
  \, \vec{e}_\vec{k}^{\lambda*} 
   \, \chi^{*}_k(t) 
   \, e^{-i \vec{k}\cdot \vec{x} }_{ }
 \,\Bigr]
 \;, \la{initial_Ai}
\ee
where the (circular) polarization vectors satisfy 
$
 \vec{k}\times \vec{e}_\vec{k}^{\pm}
 = 
 \pm i k\, \vec{e}_\vec{k}^{\pm}
$, 
and the mode function is like in \eq\nr{canonical} but 
with zero mass, 
$ \chi^{ }_k(t) \equiv e^{-i k t}_{ }/\sqrt{2 k}
$.
Via \eq\nr{eom_cont_Ai}, we then obtain an initial value 
for the electric field, similar to \eq\nr{initial_varphi}, 
\be
 \quad
 \vec{E}(0,\vec x)
  \; = \; 
 \int_\vec{k} 
 \sum_{\lambda = \pm}
 (ik)
 \Bigl[\,
  -\, c^{ }_\vec{k} 
  \, \vec{e}_\vec{k}^{\lambda} 
  \, \chi^{ }_k(0) 
  \, e^{i \vec{k}\cdot \vec{x} }_{ }
  +
   c^*_\vec{k} 
  \, \vec{e}_\vec{k}^{\lambda*} 
   \, \chi^{*}_k(0) 
   \, e^{-i \vec{k}\cdot \vec{x} }_{ }
 \,\Bigr]
 \;. 
 \quad
 \la{initial_Ei} 
\ee
For the magnetic field, in a free limit we would similarly get
\be
 \quad
 \vec{B}(0,\vec x)
  \; \overset{\rmii{free}}{\approx} \; 
 \int_\vec{k} 
 \sum_{\lambda = \pm}
 (i\vec{k}) \times 
 \Bigl[\,
  \, c^{ }_\vec{k} 
  \, \vec{e}_\vec{k}^{\lambda} 
  \, \chi^{ }_k(0) 
  \, e^{i \vec{k}\cdot \vec{x} }_{ }
  -
   c^*_\vec{k} 
  \, \vec{e}_\vec{k}^{\lambda*} 
   \, \chi^{*}_k(0) 
   \, e^{-i \vec{k}\cdot \vec{x} }_{ }
 \,\Bigr]
 \;. 
 \quad
 \la{initial_Bi} 
\ee
However, as discussed in \se\ref{ss:initial}, in our default setup we 
set $\vec{B}(0,\vec{x}) = \vec{0}$ in order to satisfy the
Gauss constraint, though a non-zero value is generated very rapidly
by the dynamics. 

\newpage

%
\section{Equations of motion on a spatial lattice}
\la{se:latt}

Following the derivation of the continuum equations 
in \app\ref{se:cont}, we transcribe them here 
to spatial discretization. 
The lattice spacing is denoted by $a$. The discretization is not 
unique; we implement it in a simple way, respecting gauge invariance, 
and comment on possibilities for ``improvement'' along the way. 

%
\subsection{Notation and canonical coordinates}
\la{ss:notation}

The basic variables of the lattice theory are 
the scalar field $\varphi(x)$, 
which lives on the sites $x$; the group elements $U^{ }_i(x)$, 
which live on the 
links between $x$ and $x + a \vec{i}$; and the temporal field component
$A^{ }_0(x)$, which lives on the sites. 
With these degrees of freedom, the $0i$-component 
of a field strength tensor 
is written in a form analogous to \eq\nr{Fmunu}, 
\be
 \widehat{F}^{ }_{0i}(x) \; \equiv \; 
 \frac{1}{iag} [\dot{U}^{ }_i(x)] U^\dagger_i(x) 
 \;+\; 
 \frac{A^{ }_0(x) 
 - U^{ }_i(x) A^{ }_0(x+a\vec{i})U^\dagger_i(x) }{a}
 \;. \la{F0i_latt}
\ee
It can be verified that under a  
gauge transformation, $\Omega(x)\in\; $SU($\Nc^{ }$), defined as   
\be
 U'_i(x) \; \equiv \; \Omega(x) U^{ }_i(x) \Omega^\dagger_{ }(x+a\vec{i})
 \;, \quad
 A'_0(x) \; \equiv \; \Omega(x) A^{ }_0(x)\Omega^\dagger_{ }(x) 
 + \frac{1}{ig} \Omega(x) \dot{\Omega}^\dagger_{ }(x)
 \;, 
\ee
the field strength transforms covariantly,  
$
 \widehat{F}^{ }_{0i}(x) 
 \to 
 \Omega(x) 
 \widehat{F}^{ }_{0i}(x) 
 \Omega^\dagger_{ }(x)
$.

The spatial components of 
the field strength are discretized by making use of a plaquette, 
\be
   P^{ }_{ij}({x}) \;\equiv\; 
   U^{ }_{i}({x}) 
   U^{ }_{j}({x} + a \vec{i})
   U^{\dagger}_{i}({x} + a \vec{j})
   U^\dagger_{j}({x}) 
   \;, \quad
   P^{ }_{ji} \; = \; P^\dagger_{ij}
   \;. \la{Pij}
\ee
In the naive continuum limit, the plaquette can be represented as  
$
 P^{ }_{ij}(x) \simeq e^{i a^2_{ }g F^{ }_{ij}}_{ }
$, 
and therefore the term appearing in the Lagrangian is defined
as 
\be
 \frac{\widehat{F^a_{ij} F^a_{ij}}}{4}
 \; \equiv \; 
 \frac{1 
 }{a^4_{ }g^2_{ }}
 \,\tr [\mathbbm{1} - P^{ }_{ij}]
 \;. 
\ee

Once we couple the axion field and the gauge fields
through the operator 
$
 \sim \kappa\hspace*{0.3mm} \varphi\, \epsilon^{ }_{ijk} F^{ }_{0i} F^{ }_{jk}
$, 
we have freedom 
with the positioning of the objects appearing. 
In a ``naive'' discretization, we make use of 
$\widehat{F}^{ }_{0i}$ from \eq\nr{F0i_latt}
as a substitute for $F^{ }_{0i}$.\footnote{%
 It has been argued that the topological charge density
 should be represented via 
 an {\em improved} discretization. 
 In the context of 
 the Euclidean topological susceptibility, a symmetrized version 
 of $\widehat{F}^{ }_{0i}$ was introduced~\cite{old},
 \be
  \langle\, \widehat{F}_{0i}\rangle (x) 
  \; \equiv \; 
  \frac{\widehat{F}_{0i}(x) + 
     U^{ }_{-i}(x) 
     \widehat{F}_{0i}(x-a\vec{i})
     U^{\dagger}_{-i}(x) 
  }{2}
  \;. \la{impr_F0i_symm}
 \ee
 Closer to our context,  
 it was seen in ref.~\cite{clgt} that the real-time 2-point
 correlator of the improved topological charge density behaves more
 regularly at small time separations than that of the naive one.
 Our study differs from these previous contexts, as it involves a dynamical
 axion field, and 2-point correlators of the topological charge 
 density do not appear directly. Nevertheless, we have also  
 worked out the equations of motion following from \eq\nr{impr_F0i_symm}. 
 In the end, we do {\em not} find any qualitative difference
 compared with the naive discretization. 
 For instance, the 
 time dependence of the Gauss constraint
 approximates to \eq\nr{dot_G_expanded} in both cases. 
 }

For the magnetic field, it is conventional to make use of a ``clover''
(cf.\ \fig\ref{fig:clover}), 
\ba
 \widehat{B}^{ }_i(x)
  & \equiv & 
 T^a_{ } \widehat{B}^{a}_i(x)
 \;, \quad
 \widehat{B}^{a}_i(x)
 \; \equiv \;
 2\,\tr\biggl\{\, T^a_{ } \biggl[\,
 \frac{  \epsilon^{ }_{ijk}
 \langle\,\widehat{F}^{ }_{jk}\,\rangle(x)
 }{2}
 \,\biggr]\,\biggr\}
 \;, \la{B_1} \\[2mm]
 \langle\, \widehat{F}^{ }_{jk}\,\rangle(x) 
 & \equiv & 
 \frac{ 
    P^{ }_{jk} + P^{ }_{k-j} + P^{ }_{-j-k} + P^{ }_{-k j}
  - P^{ }_{kj} - P^{ }_{-j k} - P^{ }_{-k-j} - P^{ }_{j -k}
 }{8 i a^2_{ }g}
 \;. \hspace*{5mm} \la{B_2}
\ea
Here the movement in a negative direction is defined 
via $U^{ }_{-i}(x) \equiv U^\dagger_{i}(x - a \vec{i})$. 
In general the components of traceless matrices are obtained as 
\be
 [...]^a_{ } \; \equiv \; 2\, \tr \{ T^a_{ } [...] \}
 \;. \la{notation} 
\ee

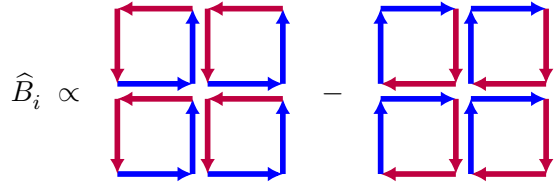
\begin{figure}[t]

\hspace*{3.5cm}%
\begin{tikzpicture}
    \node at (2.2,0){$\widehat{B}^{ }_i \;\propto \hspace*{3.1cm} - $};
    \node at (2.5,0){\clover};
    \node at (6,0){\clovercc};
\end{tikzpicture}

\caption[a]{\small
  Illustration of the clover operator, 
  defined in \eq\nr{B_2}. Red links are Hermitean conjugated, 
  while the blue ones are not. 
  }

\la{fig:clover}
\end{figure}

With this notation, the lattice Lagrangian before gauge fixing
can be written in a form analogous to \eq\nr{L_expl}, {\it viz.}\
\be
 \widehat{L} 
 \; \equiv \;
 a^3_{ }\sum_\vec{x} 
 \biggl\{ 
     \frac{1}{2} \dot{\varphi}^2_{ }
    - \frac{1}{2}  \widehat\partial^{ }_i\varphi\, 
                   \widehat\partial^{ }_i \varphi
    - V(\varphi)
    + \frac{1}{2} \widehat{F}^a_{0i} \widehat{F}^a_{0i}
    - \frac{\widehat{F^a_{ij} F^a_{ij}}}{4} 
    - \kappa\, \varphi  \,\widehat{F}^{a}_{0i} 
                        \,\widehat{B}^a_i
 \biggr\}
 \;. \la{L_latt} 
\ee
The spatial derivative acting on $\varphi$, 
denoted by $ \widehat\partial^{ }_i $, can be defined
as a difference. 

While \eq\nr{L_latt} is a perfectly well-defined gauge-invariant
Lagrangian, we have to be careful when we choose canonical coordinates.
Naively a possibility could be to parametrize the 
links through gauge potentials via
\be
 U^{ }_i(x)
 \;
 \overset{?}{=}  
 \; e^{i a g A^{ }_i(x)}_{ }
 \;. \la{link}
\ee
One problem is that then derivatives become complicated, 
\ba
      (\partial^{ }_{A^a_i} U^{ }_i)
       \, U^\dagger_i
 & = & 
 i a g\, T^a_{ }   + \frac{(iag)^2_{ }}{2} [A^{ }_i,T^a_{ }]
               + \frac{(iag)^3_{ }}{6} [A^{ }_i,[A^{ }_i,T^a_{ }]]
               + ...
 \;, \\[2mm]
 \dot{U}^{ }_i\, U^\dagger_i 
 & = & 
 i a g\, \dot{A}^{ }_i + \frac{(iag)^2_{ }}{2} [A^{ }_i,\dot{A}^{ }_i]
               + \frac{(iag)^3_{ }}{6} [A^{ }_i,[A^{ }_i,\dot{A}^{ }_i]]
               + ...  
 \;. \la{time_derivative}
\ea
It is not clear how to re-express the right-hand sides in terms of 
the $U^{ }_i$ in a useful way.

On the other hand, the right-hand side of \eq\nr{time_derivative} is
a traceless anti-Hermitean matrix. So we may define
\be
 \widehat{E}^{ }_i(x) 
 \; \equiv \; 
 \frac{1}{iag} [\dot{U}^{ }_i(x)] U^\dagger_i(x) 
 \;, \la{def_latt_Ei}
\ee
implying 
\be
 \dot{U}^{ }_i(x) 
 \; = \; 
 i a g 
 \widehat{E}^{ }_i(x) 
 U^{ }_i(x)
 \;, \quad
 \dot{U}^{\dagger}_i(x) 
 \; = \; 
 - i a g \,
 U^{\dagger}_i(x)
 \widehat{E}^{ }_i(x) 
 \;. 
\ee
The challenge is to extract the corresponding 
coordinates $\widehat{A}^a_i$, needed for 
the Euler-Lagrange equations. Fortunately,  
we only need a derivative with respect to $\widehat{A}^a_i$, 
not the values of these coordinates themselves. 
The derivative can be defined via a left action 
on the group manifold (see, e.g.,\ ref.~\cite{old2}),
\be
 \frac{\partial f [U^{ }_i(x)]}{\partial \widehat{A}^a_i(x)}
 \; \equiv \; 
 \nabla^a_{x,i} f [U^{ }_i(x)]
 \; \equiv \;
 \lim_{\epsilon\to 0}
 \frac{f\bigl[e^{i a g \epsilon T^a_{ } }_{ }\,U^{ }_i(x)\bigr]
 - f[U^{ }_i(x)] }{\epsilon}
 \;. \la{chain_rule}
\ee
This guarantees that a time derivative  
can be expressed in a usual way, 
\be
 \dot{f} 
 \; = \; 
 \sum_{\vec{x},a,i}
 \widehat{E}^a_i(x) 
 \nabla^a_{x,i} f
 \;. \la{time_der}
\ee
We adopt this implicit definition 
of the canonical coordinates in the following. 

%
\subsection{Gauss constraint and Hamiltonian}
\la{ss:setup}

When we fix to the gauge $A^a_0 = 0$, the Gauss constraint needs to be
imposed as a consistency relation. 
Let us derive its discretized version. In analogy
with \eq\nr{Gauss_cont_pre}, it reads
\be
  0 \; = \; \widehat{G}^{ }_{ }(x) 
 \; \equiv \;  
 T^a_{ }\frac{\delta \widehat{L}}{\delta A^a_0(x)} \bigg|^{ }_{A^a_0 = 0}
 \;, \la{Gauss_def_latt}
\ee
where we have defined
$
 {\delta \widehat{L}} / {\delta A^a_0}
 \equiv
 (1/a^3_{ }) 
 {\partial \widehat{L}} / {\partial A^a_0}
$.
We discuss separately the contributions
present at $\kappa = 0$, and the corrections at first order in $\kappa$, 
writing
\be
 \widehat{G}(x) \;=\;
 \widehat{G}(x) \bigr|^{ }_{\kappa = 0}
 + \widehat{G}(x) \bigr|^{ }_{\rmO(\kappa)}
 \;. \la{Gauss_both}
\ee 
At zeroth order, from \eqs\nr{F0i_latt} and \nr{L_latt}, 
adopting the definition in \eq\nr{def_latt_Ei}, 
\be
 \widehat{G}(x) \bigr|^{ }_{\kappa = 0}
 \; 
 \underset{\rmii{\nr{def_latt_Ei},\nr{Gauss_def_latt}}}{
 \overset{\rmii{\nr{F0i_latt},\nr{L_latt}}}{=}} 
 \; 
 \frac{1}{a}
 \sum_i
 \bigl[\,
  \widehat{E}^{ }_i(x) 
  - 
  U^{ }_{-i}(x) 
  \widehat{E}^{ }_i(x-a\vec{i})
  U^{\dagger}_{-i}(x) 
 \,\bigr]
 \;. 
 \la{Gauss_kappa0}
\ee
At first order, 
\be
 \widehat{G}(x) \bigr|^{ }_{\rmO(\kappa)}
 \;
 \underset{\rmii{\nr{def_latt_Ei},\nr{Gauss_def_latt}}}{
 \overset{\rmii{\nr{F0i_latt},\nr{L_latt}}}{=}} 
 \; 
 \frac{\kappa}{a}
 \sum_i
 \bigl[\,
  U^{ }_{-i}(x) 
  \varphi(x-a\vec{i})
  \widehat{B}^{ }_i(x-a\vec{i})
  U^{\dagger}_{-i}(x) 
   - 
  \varphi(x)
  \widehat{B}^{ }_i(x)
 \,\bigr]
 \;. 
 \la{Gauss_kappa1}
\ee
This generalizes the second term of \eq\nr{Gauss_cont_pre}.

The Hamiltonian is obtained from a Legendre transform like 
in \eq\nr{Legendre}. 
By construction, terms depending
linearly on $\widehat{E}^a_i$ cancel in the Legendre transform, 
which then leads to 
\be
 \widehat{H}
 \;
 \underset{\rmii{\nr{Legendre}}}{
 \overset{\rmii{\nr{L_latt}}}{=}}
 \;
 a^3_{ }
 \sum_\vec{x} 
 \biggl\{ 
     \frac{1}{2} \dot{\varphi}^2_{ }
    + \frac{1}{2}  \widehat{\partial}^{ }_i\varphi
                \, \widehat{\partial}^{ }_i \varphi
    + V(\varphi)
    + \frac{1}{2} \widehat{E}^a_{i}\widehat{E}^a_{i}
    + \frac{1}{a^4_{ }g^2_{ }}
      \sum_{i,j} \tr [\mathbbm{1} - P^{ }_{ij}] 
 \biggr\}
 \;. \la{H_latt}
\ee

%
\subsection{Evolution equations}
\la{ss:eom}

The Euler-Lagrange equations are obtained like in \eq\nr{E-L}. 
On the left-hand side, 
\be
 \frac{\delta \widehat{L}}{\delta \widehat{E}^a_i(x)} \bigg|^{ }_{A^a_0 = 0}
 \;
 \underset{\rmii{\nr{chain_rule}}}{
 \overset{\rmii{\nr{L_latt}}}{=}}
 \; 
 \widehat{E}^a_i(x) 
 - {\kappa} \,
  \varphi(x) \widehat{B}^{a}_i(x)
 \;. \la{dEai_hatL}
\ee

In order to work out 
the right-hand side, we introduce the short-hand notation
\be
 Q^{ }_{mn...}(y) f
 \; \equiv \; 
 U^{ }_m(y)\, U^{ }_n(y+a\vec{m}) ... f(y + a\vec{m} + a \vec{n} + ...)
 \;, \la{def_Q}
\ee
which implies $U^{ }_i = Q^{ }_i$ and
$P^{ }_{ij} = Q^{ }_{ij-i-j}$.
Then, for a single plaquette, \eq\nr{chain_rule} yields
\ba
 \frac{\partial P^{ }_{jk}(y)}{\partial \widehat{A}^a_i(x)}
 &
 \underset{\rmii{\nr{chain_rule}}}{
 \overset{\rmii{\nr{Pij}}}{=}}
 & 
 i a g\, \bigl\{\,
 \delta^{ }_{i,j} 
 \bigl[\,
 \delta^{ }_{\vec{x},\vec{y}} \, T^a_{ } Q^{ }_{jk-j-k}(x)
 - 
 \delta^{ }_{\vec{x},\vec{y}+a\vec{k}}
   \,Q^{ }_{jk-j}(y)\, T^a_{ }\, Q^{ }_{-k}(x)
 \,\bigr]
 \nn[2mm]
 & & \hspace*{5mm}
 + \,\delta^{ }_{i,k} 
  \bigl[\,
   \delta^{ }_{\vec{x},\vec{y}+a\vec{j}}
   \,Q^{ }_j(y)\, T^a_{ } Q^{ }_{k-j-k}(x)
  - 
  \delta^{ }_{\vec{x},\vec{y}} Q^{ }_{jk-j-k}(x) T^a_{ }
  \,\bigr]
 \,\bigr\}
 \;. \hspace*{8mm} \la{dAai_Pjk}
\ea
After taking a trace, this leads 
to the first line of \eq\nr{eom_latt_Eai}. 

In contrast, 
the terms of $\rmO(\kappa)$ are cumbersome. Moving the second term
from \eq\nr{dEai_hatL} on the right-hand side, we get
\be
 \dot{\widehat{E}}\hspace*{0.4mm}{}^a_i(x) \bigr|^{ }_{\rmO(\kappa)}
 \;
 \overset{\rmii{\nr{E-L}}}{
 \underset{\rmii{\nr{L_latt},\nr{dEai_hatL}}}{=}} 
 \; 
 {\kappa}\,
 \Bigl\{\, 
  \dot{\varphi}(x) \widehat{B}^a_i(x)
 + 
   \varphi(x) \dot{\widehat{B}}\hspace*{0.4mm}{}^a_i(x)
  -  
 \sum_{\vec{y},c,j} \widehat{E}^c_j(y)
  \varphi(y) 
  \nabla^a_{x,i}
  \widehat{B}^c_j(y)
 \,\Bigr\}
 \;. \la{dot_Eai_kappa1}
\ee

The time derivative of the magnetic field on 
the second line of \eq\nr{dot_Eai_kappa1} can be taken with
\eq\nr{time_der}. After some algebra we obtain, 
employing the notation from \eq\nr{notation},
\be
 \dot{\widehat{B}}\hspace*{0.4mm}{}^a_i(x) = 
 \frac{\epsilon^{ }_{ijk}\,
 \{ \Phi^{ }_{jk}(x) \}^a_{ }
 }{8a}
 \;, \la{def_Phi}
\ee
where, making use of the notation in \eq\nr{def_Q},
\ba
 \Phi^{ }_{jk} & = & 
   \widehat{E}^{ }_j\, Q^{ }_{jk-j-k}
 + \widehat{E}^{ }_k\, Q^{ }_{k-j-kj}
 - Q^{ }_{-kjk-j}\,\widehat{E}^{ }_j
 - Q^{ }_{jk-j-k}\,\widehat{E}^{ }_k 
 \nn[2mm]
 & + & 
   Q^{ }_{-k}\widehat{E}^{ }_j Q^{ }_{jk-j}
 + Q^{ }_{j}\widehat{E}^{ }_k Q^{ }_{k-j-k}
 - Q^{ }_{-j} \widehat{E}^{ }_j Q^{ }_{-kjk}
 - Q^{ }_{-k} \widehat{E}^{ }_k Q^{ }_{jk-j}
 \nn[2mm]
 & + & 
   Q^{ }_{-j-k} \widehat{E}^{ }_{j} Q^{ }_{jk}
 + Q^{ }_{-kj} \widehat{E}^{ }_{k} Q^{ }_{k-j}
 - Q^{ }_{k-j} \widehat{E}^{ }_{j} Q^{ }_{-kj}
 - Q^{ }_{-j-k} \widehat{E}^{ }_{k} Q^{ }_{jk}
 \nn[2mm]
 & + & 
   Q^{ }_{k-j-k} \widehat{E}^{ }_{j} Q^{ }_j
 + Q^{ }_{-j-kj} \widehat{E}^{ }_{k} Q^{ }_k
 - Q^{ }_{jk-j} \widehat{E}^{ }_j Q^{ }_{-k}
 - Q^{ }_{k-j-k} \widehat{E}^{ }_k Q^{ }_j
 \;. \la{Phi_jk}
\ea
From the last term of \eq\nr{dot_Eai_kappa1}, we get 
\be
   -  
 \sum_{\vec{y},c,j} 
 \widehat{E}^{c}_j(y)
  \varphi(y)
  \nabla^a_{x,i}
  \widehat{B}^c_j(y)
 \; = \; 
 \frac{\epsilon^{ }_{ijk}\,
 \{ \Theta^{ }_{ijk}(x)  \}^a_{ }
 }{8a}
 \;, \la{Theta_origin}
\ee
where
\ba
 \Theta^{ }_{ijk} 
 & = & 
   \varphi \naive{E}{j} Q^{ }_{ki-k-i}
 - \varphi \naive{E}{j} Q^{ }_{-kik-i}
 + Q^{ }_{ik-i-k} \varphi\naive{E}{j}
 - Q^{ }_{i-k-ik} \varphi\naive{E}{j}
 \nn[2mm]
 & + & 
   Q^{ }_k \varphi\naive{E}{j} Q^{ }_{i-k-i}
 + Q^{ }_i \varphi\naive{E}{j} Q^{ }_{k-i-k}
 - Q^{ }_{-k} \varphi\naive{E}{j} Q^{ }_{ik-i}
 - Q^{ }_{i} \varphi\naive{E}{j} Q^{ }_{-k-ik}
 \nn[2mm]
 & + & 
   Q^{ }_{ki} \varphi\naive{E}{j} Q^{ }_{-k-i}
 + Q^{ }_{ik} \varphi\naive{E}{j} Q^{ }_{-i-k}
 - Q^{ }_{-ki} \varphi\naive{E}{j} Q^{ }_{k-i}
 - Q^{ }_{i-k} \varphi\naive{E}{j} Q^{ }_{-ik}
 \nn[2mm] 
 & + & 
   Q^{ }_{ki-k} \varphi\naive{E}{j} Q^{ }_{-i}
 + Q^{ }_{ik-i} \varphi\naive{E}{j} Q^{ }_{-k}
 - Q^{ }_{i-k-i} \varphi\naive{E}{j} Q^{ }_{k}
 - Q^{ }_{-kik} \varphi\naive{E}{j} Q^{ }_{-i}
 \;. \hspace*{6mm} \la{Theta_ijk}
\ea

In a practical implementation, it is important to 
minimize matrix multiplications. In particular, 
\eq\nr{Theta_ijk} can 
be regrouped in various ways for this purpose, for instance
\ba
 \Theta^{ }_{ijk} 
 & \overset{\rmii{\nr{Theta_ijk}}}{=} & 
   \bigl(\,
              \varphi \naive{E}{j} Q^{ }_{ki}
 + Q^{ }_k    \varphi \naive{E}{j} Q^{ }_i 
 + Q^{ }_{ki} \varphi \naive{E}{j}
   \,\bigr) \, Q^{ }_{-k-i}
 \nn[2mm]
 & - & 
   \bigl(\,
               \varphi \naive{E}{j} Q^{ }_{-ki}
 + Q^{ }_{-k}  \varphi \naive{E}{j} Q^{ }_i 
 + Q^{ }_{-ki} \varphi \naive{E}{j}
   \,\bigr) \, Q^{ }_{k-i}
 \nn[2mm]
 & + & 
 Q^{ }_i \varphi\naive{E}{j}
 \,\bigl(\,
    Q^{ }_{k-i-k} - Q^{ }_{-k-ik}
 \,\bigr)
 \; + \; \mbox{H.c.}
 \;, \hspace*{6mm} \la{Theta_ijk_simpl}
\ea
where ``H.c.''\ denotes Hermitean conjugation of the whole expression. 

\vspace*{3mm}

Putting everything together, 
the analogues of \eqs\nr{eom_cont_scalar}--\nr{eom_cont_Ei}
can be expressed as 
\ba
 \ddot{\varphi}(x)  & = & 
 \widehat{\Delta}\,
 \varphi(x)
 - V'(\varphi)
 - \kappa 
  {\textstyle\sum_{a,i}} 
   \widehat{E}^a_i(x)
   \widehat{B}^a_i(x)
 \;, \la{eom_latt_scalar} \\[2mm]
 \dot{U}^{ }_i(x) & = & i a g \, \widehat{E}^{ }_i(x) U^{ }_i(x)
 \;, \la{eom_latt_Ai} \\[2mm]
 \dot{\widehat{E}}\hspace*{0.4mm}{}^a_i(x) 
 & = & 
 \frac{i}{a^3_{ }g}
 {\textstyle\sum_j}
 \tr \bigl\{ T^a_{ } 
 \bigl[\,
   P^{ }_{ij} + P^{ }_{i-j} - P^{ }_{ji} - P^{ }_{-ji}
 \,\bigr](x)
 \bigr\}
 \nn[2mm] 
 & + & 
 \kappa \,  
 \biggl\{\, 
  \dot{\varphi}(x) \widehat{B}^{ }_i(x)
 + 
 {\textstyle\sum_{j,k}}
 \frac{\epsilon^{ }_{ijk}\,
 \bigl[\,
 \varphi(x)\Phi^{ }_{jk}(x) 
 + \Theta^{ }_{ijk}(x) 
 \,\bigr] 
 }{8a}
 \biggr\}^a_{ }
 \;,  \la{eom_latt_Eai}  
\ea
where $  \widehat{\Delta} $ is  a lattice Laplacian. 
In comparison with \eq\nr{eom_cont_Ei}, 
the first row of \eq\nr{eom_latt_Eai} represents 
$
 - 
  \mathcal{D}\times \vec{B}
$, 
the terms on the second row
$
  \kappa \, 
   \dot\varphi\, \vec{B}
$
and 
$
 - \kappa \nabla\varphi \times \vec{E}
$, 
respectively.

Even though \eqs\nr{eom_latt_scalar}--\nr{eom_latt_Eai} constitute
the perfect analogue of the continuum equations in 
\eqs\nr{eom_cont_scalar}--\nr{eom_cont_Ei}, we find it
efficient to reorganize them. Specifically, let us define
\be
 \widehat\Pi^a_i(x) 
 \; \equiv \; 
 \widehat{E}^a_i(x)
 - \kappa \varphi(x) \widehat{B}^a_i(x)
 \;. \la{def_Pi}
\ee
It follows from \eqs\nr{dot_Eai_kappa1} and \nr{def_Phi} that
if we compute the time evolution of $\widehat\Pi^a_i$, 
then the complicated object $\Phi^{ }_{jk}$, given in \eq\nr{Phi_jk}, 
does not need to be evaluated. 

Furthermore, we rescale variables and parameters 
into dimensionless forms. For this we make use of 
the physical length scale, $\scale$, 
introduced in \eq\nr{rescale0}. In principle it could be  
associated for instance with the inverse Hubble rate
during inflation, $H^{-1}_{ }$, or the inverse mass, $m^{-1}_{ }$. 
In the absence of expansion, we follow the 
latter possibility (cf.\ \eq\nr{rescale0}). 
All dimensionful quantities
(fields, coordinates, parameters) are scaled with
$\scale$ so as to make them dimensionless, 
\ba
 \widetilde\varphi & \equiv & \scale\,\varphi
 \;, \quad
 \widetilde E^a_i \; \equiv \; \scale^2_{ }\, \widehat E^a_i
 \;, \quad
 \widetilde B^a_i \; \equiv \; \scale^2_{ }\, \widehat B^a_i
 \;, \quad
 \widetilde \Theta^a_{ijk} \; \equiv \; \scale^3_{ }\, \Theta^a_{ijk}
 \;, \la{rescale1} \\[3mm]
 \widetilde t & \equiv & t / \scale
 \;, \quad
 \widetilde a \; \equiv \; a / \scale
 \;, \quad
 \widetilde k \; \equiv \; \scale\, k
 \;, \quad
 \widetilde \omega^{ }_k \; \equiv \; \scale\, \omega^{ }_k
 \;, \la{rescale2} \\[3mm]
 \widetilde \kappa & \equiv & \kappa / \scale
 \;, \quad
 \widetilde m \; \equiv \; \scale\, m
 \;. \la{rescale3}
\ea
The gauge coupling, $g^2_{ }$, is dimensionless
and does not get scaled in the classical theory 
(it should be thought of as a renormalized coupling, 
$g^2_{ }(\bmu \sim 1/\scale)$).

Being also explicit about the lattice Laplacian, and inserting
the potential from \eq\nr{V}, 
it follows from \eqs\nr{eom_latt_scalar}--\nr{rescale3} that 
\begin{empheq}[box=\fbox]{align}
 \partial^2_{\tilde t}\widetilde\varphi(x)  
 & \;=\;  
  {\textstyle\sum_{i}} 
 \frac{ 
 \widetilde\varphi(x + a\vec{i})
 - 2 \widetilde\varphi(x) 
 + \widetilde\varphi(x - a\vec{i})
 }{\tilde a^2_{ }}
 - \widetilde m^2_{ }\widetilde \varphi
 - \widetilde \kappa\, 
  {\textstyle\sum_{a,i}} 
   \widetilde{E}^a_i(x)
   \overbrace{ \widetilde{B}^a_i(x) }^{\rmii{\nr{B_1}}}
 \;, \la{eom_latt_scalar_resc} \vphantom{\bigg | } \\[3mm]
 \partial^{ }_{\tilde t }{U}^{ }_i(x) 
 & \;=\; i \tilde a g \widetilde{E}^{ }_i(x) U^{ }_i(x)
 \;,
 \la{eom_latt_Ai_resc}
 \\[-1mm]
 \;
 \partial^{ }_{\tilde t}\widetilde{\Pi}\hspace*{0.4mm}{}^a_i(x) 
 &
 \; 
 \underset{\rmii{\nr{def_Phi}}}{
 \overset{\rmii{\nr{eom_latt_Eai}}}{=}}
 \;
 {\textstyle\sum_j}
 \frac{i\,
 \tr \bigl\{ T^a_{ } 
 \bigl[\,
   P^{ }_{ij} + P^{ }_{i-j} - P^{ }_{ji} - P^{ }_{-ji}
 \,\bigr](x)
 \bigr\}
 }{\tilde a^3_{ }g}
  + 
 {\textstyle\sum_{j,k}}
 \frac{\widetilde \kappa \,\epsilon^{ }_{ijk}\,
 \overbrace{
 \widetilde\Theta^{a}_{ijk}(x) }^{\rmii{\nr{Theta_ijk_simpl}}}
 }{8\tilde a}
 \;,  \;
 \la{eom_latt_Piai_resc}   
 \\[3mm]
 \widetilde{E}^a_i
 & \; \overset{\rmii{\nr{def_Pi}}}{=} \; 
 \widetilde\Pi^a_i 
 + \widetilde\kappa\, \tilde\varphi\, \widetilde{B}^a_i
 \;. \vphantom{\bigg | }
 \la{eom_Ei_resc}
\end{empheq}
What this means is that we evolve the scalar field
with \eq\nr{eom_latt_scalar_resc},
the links with 
\eq\nr{eom_latt_Ai_resc}, and
$\widetilde{\Pi}^a_i$ with \eq\nr{eom_latt_Piai_resc}.
For the next time step,  
we solve for $\widetilde{E}^a_i$ from \eq\nr{eom_Ei_resc}. 

As an example of a physical quantity, we note from \eq\nr{H_latt}
that the energy density associated with the electric fields is
$e \equiv E/V \supset \frac{1}{2} \widehat{E}^a_{i}\widehat{E}^a_{i}$.
In units of $\scale$, this becomes 
$\scale^4_{ }e \supset \frac{1}{2} \widetilde{E}^a_{i}\widetilde{E}^a_{i}$.
When we represent this in terms of 
a power spectrum (cf.\ \se\ref{ss:initial2}), the ultraviolet
(Rayleigh-Jeans) problem of classical field theory 
lies in the fact that in the course of time, much of the energy density
transitions to momenta $k\sim 1/a$, i.e.\ 
$\tilde k \sim 1/\tilde a \gg 1$
(cf.\ \se\ref{ss:equi}).

%
\subsection{Energy conservation}
\la{ss:energy}

Consider the time derivative of the 
Hamiltonian from \eq\nr{H_latt},
\be
 \dot{\widehat{H}}
 \; \overset{\rmii{\nr{H_latt}}}{=} \;
 a^3_{ }
 \sum_\vec{x} 
 \biggl\{ 
       \dot{\varphi} \ddot{\varphi}
     - \dot{\varphi}\, 
       \widehat{\Delta}  \varphi
    + \dot{\varphi} V'(\varphi)
    + \widehat{E}^a_{i}\dot{\widehat{E}}\hspace*{0.4mm}{}^a_{i}
    - \frac{1}{a^4_{ }g^2_{ }}
      \sum_{i,j} \tr [\dot{P}^{ }_{ij}] 
 \biggr\}
 \;. \la{dot_H_latt}
\ee
At order $\kappa = 0$, the scalar field parts cancel directly
thanks to \eq\nr{eom_latt_scalar}.
For the electric field part at $\kappa = 0$, 
the first row of 
\eq\nr{eom_latt_Eai} yields 
\be
 \dot{\widehat{H}} \bigr|^{ }_{\kappa\, =\, 0}
 \; 
 \underset{\rmii{\nr{eom_latt_Eai}}}{
 \overset{\rmii{\nr{dot_H_latt}}}{\supset}}
 \;
 \frac{i}{g}
 \sum_{\vec{x},i,j} \tr \bigl\{ \widehat{E}^{ }_i(x)
 \bigl[\,
   P^{ }_{ij} + P^{ }_{i-j} - P^{ }_{ji} - P^{ }_{-ji}
 \,\bigr](x)
 \bigr\}
 \;. \la{part_H_1}
\ee
From the plaquette in the last term of \eq\nr{dot_H_latt}, we obtain
\be
 \dot{\widehat{H}} \bigr|^{ }_{\kappa\, =\, 0}
 \;
  \underset{\rmii{\nr{dAai_Pjk}}}
  {\overset{\rmii{\nr{time_der}}}{ \supset }} 
 \;
 - \frac{i}{g} \sum_{\vec{x},i,j}
  \tr\bigl\{ 
    \widehat{E}^{ }_i(x) P^{ }_{ij}(x)
  -  P^{ }_{-ji}(x) \widehat{E}^{ }_i(x)
  + \widehat{E}^{ }_i(x) P^{ }_{i-j}(x)
  - P^{ }_{ji}(x) \widehat{E}^{ }_i(x) 
  \bigr\}
 \;. \la{part_H_2}
\ee
Using the cyclic property of the trace, 
\eqs\nr{part_H_1} and \nr{part_H_2} cancel 
against each other. 

Proceeding to $\rmO(\kappa)$, the last term of \eq\nr{eom_latt_scalar}
and the second term of \eq\nr{eom_latt_Eai} yield
\ba
 \dot{\widehat{H}} \bigr|^{ }_{\rmO(\kappa)}
 &
 \underset{\rmii{\nr{eom_latt_scalar},\nr{eom_latt_Eai}}}{
 \overset{\rmii{\nr{dot_H_latt}}}{\supset}}
 &
 \kappa a^3_{ } \sum_{\vec{x},c,i} 
 \Bigl\{ 
 - \dot{\varphi}(x) 
      \widehat{E}^{c}_i(x) 
      \widehat{B}^c_i(x)
 +\,
 \widehat{E}^c_i(x)
  \dot{\varphi}(x) \widehat{B}^{c}_i(x)
 \Bigr\}
 \;, 
\ea
which clearly cancel.
Finally, for the contributions originating from
the third term \linebreak of \eq\nr{eom_latt_Eai}, the terms from 
$\Phi^{ }_{jk}$ and $\Theta^{ }_{ijk}$ both give structures
containing two electric fields, 
connected by $Q$'s. 
Shifting~$\vec{x}$, re-ordering terms inside the trace, 
and renaming indices, all 16 terms find a counterpart
on the other side, and cancel exactly. 

%
\subsection{Time dependence of the Gauss constraint}
\la{ss:gauss}

Turning to 
the Gauss constraint 
from \eq\nr{Gauss_both}, the goal is to see whether it is
time-independent.  
Let us first inspect its $\kappa = 0$ part, given
by \eq\nr{Gauss_kappa0}. 
We note that two of the terms
from $\dot{\widehat{G}}$ immediately
cancel as a consequence of
\eq\nr{eom_latt_Ai}, 
\be
  - \hspace*{-6mm}
  \underbrace{
  \dot{U}^\dagger_i(x-a\vec{i})
  }_{
  -i a g U^\dagger_i(x-a\vec{i})\widehat{E}^{ }_i(x-a\vec{i})
  }
  \hspace*{-6mm}
  \widehat{E}^{ }_i(x-a\vec{i})
  U^{ }_i(x-a\vec{i})
  - 
  U^\dagger_i(x-a\vec{i})
  \widehat{E}^{ }_i(x-a\vec{i})
  \hspace*{-6mm}
  \underbrace{
  \dot{U}^{ }_i(x-a\vec{i})
  }_{
  +i a g \widehat{E}^{ }_i(x-a\vec{i}) U^{ }_i(x-a\vec{i})
  }
  \hspace*{-6mm}
 = 0 
 \;. 
\ee
Therefore 
\be
 \dot{\widehat{G}}(x) \bigr|^{ }_{\kappa = 0}
 \; \overset{\rmii{\nr{Gauss_kappa0}}}{=} \; 
 \frac{1}{a}
 \sum_i
 \Bigl[\,
  \dot{\widehat{E}}^{ }_i(x) \Bigr|^{ }_{\kappa = 0}
  - 
  U^{ }_{-i}(x) 
  \dot{\widehat{E}}^{ }_i(x-a\vec{i}) \Bigr|^{ }_{\kappa = 0}
  U^{\dagger}_{-i}(x) 
 \,\Bigr]
 \;. 
 \la{dot_Gauss_kappa0}
\ee
The time derivatives are given by the first 
row of \eq\nr{eom_latt_Eai}. 
When we sum over 
the generators $T^a_{ }$ to obtain
$\dot{\widehat{E}}^{ }_i = T^a_{ } \dot{\widehat{E}}{}^a_i$, 
then the result has the form 
\be
 \sum_a T^a_{mn} \tr \{ T^a_{ } M \}
 \; = \;  
    \underbrace{ 
    \sum_a 
    T^a_{mn} T^a_{kl } 
    }_{
    \frac{1}{2}\bigl( \delta^{ }_{ml}\delta^{ }_{nk}
                  - \frac{1}{\Nc^{ }}
                    \,\delta^{ }_{mn}\delta^{ }_{kl} \bigr)
    }
   \hspace*{-6mm}
    M^{ }_{lk}
 \; = \;
 \frac{1}{2} \biggl(
 M^{ }_{mn} - \frac{\delta^{ }_{mn} \tr M }{\Nc^{ }}
  \biggr)
 \;. 
\ee
It is enough to consider the first term, as the second 
term just subtracts its trace part, which vanishes if the 
term itself vanishes. 
Then the second term in \eq\nr{dot_Gauss_kappa0} re-orders
the indices of the plaquettes, and we get
\be
 \dot{\widehat{G}}(x) \bigr|^{ }_{\kappa = 0}
 \; 
 \underset{\rmii{\nr{dot_Gauss_kappa0}}}
 {\overset{\rmii{\nr{eom_latt_Eai}}}{\supset}}
 \; 
 \frac{i}{2a^4_{ }g}
 {\textstyle\sum_{ij}}
 \{ 
     P^{ }_{ij} + P^{ }_{i-j} - P^{ }_{ji} - P^{ }_{-ji}
   - P^{ }_{j-i} - P^{ }_{-j-i} + P^{ }_{-ij} + P^{ }_{-i-j}
 \}
 \;. 
\ee
The summand is antisymmetric in $i\leftrightarrow j$, and therefore
the sum vanishes.

Proceeding to the part of $\rmO(\kappa)$, \eqs\nr{Gauss_kappa0}
and \nr{Gauss_kappa1} yield 
\ba
 \dot{\widehat{G}}(x) \bigr|^{ }_{\rmO(\kappa)}
 &  
 \underset{\rmii{\nr{Gauss_kappa1}}}{
 \overset{\rmii{\nr{Gauss_kappa0}}}{=}}
 &
 \frac{1}{a}
 \sum_i
 \Bigl\{\,
  \dot{\widehat{E}}^{ }_i(x) \bigr|^{ }_{\rmO(\kappa)} 
  - 
  U^{ }_{-i}(x) 
  \dot{\widehat{E}}^{ }_i(x-a\vec{i})  \bigr|^{ }_{\rmO(\kappa)}
  U^{\dagger}_{-i}(x) 
 \,\Bigr\}
 \nn[2mm]
 & & \; + \,  
 \frac{\kappa}{a}
 \sum_i
 \Bigl\{ \,
  \dot\varphi(x-a\vec{i})
  U^{ }_{-i}(x) 
  \widehat{B}^{ }_i(x-a\vec{i})
  U^{\dagger}_{-i}(x) 
  - \,
  \dot\varphi(x)
  \widehat{B}^{ }_i(x)
 \nn[2mm]
 & & \hspace*{1.1cm} + \,  
  \varphi(x-a\vec{i})
  \partial^{ }_t 
  \bigl[\,
  U^{ }_{-i}(x) 
  \widehat{B}^{ }_i(x-a\vec{i})
  U^{\dagger}_{-i}(x) 
  \,\bigr]
    - \,
  \varphi(x)
  \dot{\widehat{B}}^{ }_i(x)
 \,\Bigr\}
 \;. \hspace*{5mm}
 \la{dot_Gauss_kappa1}
\ea
Inserting the second term of \eq\nr{eom_latt_Eai}
for the first row of \eq\nr{dot_Gauss_kappa1}, the resulting 
terms cancel against
the second row of \eq\nr{dot_Gauss_kappa1}. 

It remains to combine the third term of \eq\nr{eom_latt_Eai}
with the third row of \eq\nr{dot_Gauss_kappa1}.  
The terms containing $\Phi^{ }_{jk}$ cancel exactly against 
the time derivatives of $\widehat{B}^{ }_i$ from 
the third row of \eq\nr{dot_Gauss_kappa1}.
For the terms containing $\Theta^{ }_{ijk}$, we find an almost
complete cancellation after renaming indices and making use of
antisymmetry, leaving over just
\ba
  \dot{\widehat{G}}(x) \bigr|^{ }_{\rmO(\kappa)}
 & \supset & \frac{\kappa}{8 a^2_{ }}\sum_{i,j,k}
 \epsilon^{ }_{ijk}
 \,\bigl\{\,
     \Theta^{ }_{ijk}(x) 
    - 
    U^{ }_{-i}(x) 
    \Theta^{ }_{ijk}(x-a\vec{i})
    U^{\dagger}_{-i}(x) 
 \,\bigr\}
 \nn[2mm]
 & = & 
 \frac{\kappa}{8 a^2_{ }}\sum_{i,j,k}
 \epsilon^{ }_{ijk}
 \bigl[\,
   \varphi \naive{E}{i} \,,\,
   P^{ }_{jk} + P^{ }_{k-j} + P^{ }_{-j-k} + P^{ }_{-kj}
 \,\bigr] 
 \;. \la{commutator}
\ea

Commutators similar to 
\eq\nr{commutator} arise also from when $\partial^{ }_t$
acts on $U^{ }_i$ or $U^\dagger_i$ on 
the third row of \eq\nr{dot_Gauss_kappa1}. 
Combining the latter terms with \eq\nr{commutator}, 
and re-expressing the result
in terms of the magnetic field from 
\eqs\nr{B_1} and \nr{B_2}, we find the full result
\ba
 \dot{\widehat{G}}(x) 
 & = &  
 i g \kappa 
 \sum_i 
 \Bigl\{\,
 \varphi(x)
 \bigl[\,
 \widehat{E}^{ }_i(x)
 \,,\,  
 \widehat{B}^{ }_i(x)
 \,\bigr]
 \nn[2mm]
 & & \; - \,  
 U^{ }_{-i}(x) 
 \varphi(x - a \vec{i})
 \bigl[\,
 \widehat{E}^{ }_i(x -  a \vec{i})
 \,,\, 
 \widehat{B}^{ }_i(x - a \vec{i})
 \,\bigr] 
 U^{\dagger}_{-i}(x) 
 \,\Bigr\}
 \;. 
 \la{dotG_improved}
\ea

In general, \eq\nr{dotG_improved} is non-vanishing. 
However, because it contains commutators, 
it vanishes in an Abelian theory, and because 
it is composed of a difference of two similar terms, 
it vanishes when the lattice 
spacing $a$ is small compared with the dynamical scales
of the system ($a \ll \scale$). To be specific 
about the latter, a formal expansion in $a$ gives
\be
 \dot{\widehat{G}}(x) 
 \; \simeq \;  
  i g \kappa 
 \sum_i \bigl\{\, 
 a \bigl[\, D^{ }_i , \varphi(x)  \bigl[\, E^{ }_i(x), B^{ }_i(x) 
 \,\bigr]\,\bigr] 
 + \rmO(a^2_{ })
 \bigr\} 
 \;. \la{dot_G_expanded}
\ee
For the lattice spacing that we have
used in practice 
(cf.\ \eq\nr{def_tilde_a}), this time
evolution is numerically slow, but
nevertheless
clearly observable in the SU(2) case.

It is unclear to us whether the existence of 
\eq\nr{dot_G_expanded}
has a deeper reason, related to the non-topological 
nature of the lattice discretization of the topological charge
density. From the simulation point of view, it however
represents no practical challenge. 

%
\subsection{Initial conditions and power spectra}
\la{ss:initial2}

When we study the system on a lattice, the initial conditions 
in \eqs\nr{initial_varphi}--\nr{variance_new} need to be rewritten. 
With a finite lattice spacing, $a$, and a finite lattice
extent, $N$, across which period boundary conditions are imposed, 
momenta, integrals, and $\delta$-constraints take the forms
\ba
 \vec{k}
 &
 \to
 & 
 \vec{k}(\vec{n})
 \; \equiv \; 
 \frac{2\pi\vec{n}}{a N}
 \;, \quad 
 \vec{n} 
 \;
 \in 
 \; 
 \Bigl[ -\frac{N}{2},...,\frac{N}{2} - 1 \Bigr]^3_{ }
 \;, \la{k_lat}
 \\[3mm]
 \int_\vec{k} 
 & 
  \underset{\rmii{\nr{k_lat}}}{
  \overset{\rmii{\nr{mode_expansion}}}{\to}}
 & 
 \frac{1}{a^3_{ }N^3_{ }}
 \sum_\vec{n} 
 \;, \quad
 \langle \, c^{ }_\vec{n} c^*_\vec{m} \, \rangle
 \; 
 \overset{\rmii{\nr{variance}}}{\to}
 \; 
 f^{ }_{k}\, a^3_{ }N^3_{ } \, 
 \delta^{ }_{\vec{n},\vec{m}\;\rmi{mod}\;\vec{N}}
 \;. \la{variance_lat}
\ea
It is convenient to normalize the noise close to unity, so we rescale
$
 c^{ }_\vec{n} \to \sqrt{a^3_{ }N^3_{ }}\, \hat c^{ }_\vec{n}
$.
Furthermore, we rescale fields and momenta according to 
\eqs\nr{rescale1} and \nr{rescale2}, respectively. 
Then \eqs\nr{initial_varphi} and \nr{initial_dvarphi} become
\begin{align}
 \quad
 \widetilde \varphi(0,\vec x)
 & \; = \; 
 \frac{1}{\sqrt{\tilde a^3_{ }N^3_{ }}} 
 \sum_\vec{n}
 \bigl[\,
   \hat c^{ }_\vec{n} \, \widetilde\varphi^{ }_k(0) 
    \, e^{i \vec{k}(\vec{n})\cdot \vec{x} }_{ }
 + 
   \hat c^*_\vec{n} \, \widetilde\varphi^{*}_k(0) 
    \, e^{-i \vec{k}(\vec{n})\cdot \vec{x} }_{ }
 \,\bigr]
 \;, 
 \la{initial_varphi_lat}
 \\[3mm]
 \partial^{ }_{\tilde t}\hspace*{0.3mm}\widetilde\varphi(0,\vec x)
 & \; = \; 
 \frac{1}{\sqrt{\tilde a^3_{ }N^3_{ }}} 
 \sum_\vec{n} (i \widetilde\omega^{ }_k)
 \bigl[\,
   -\, \hat c^{ }_\vec{n} \, \widetilde\varphi^{ }_k(0)
    \, e^{i \vec{k}(\vec{n})\cdot \vec{x} }_{ }
  +
   \hat c^*_\vec{n} \, \widetilde\varphi^{*}_k(0) 
   \, e^{-i \vec{k}(\vec{n})\cdot \vec{x} }_{ }
 \,\bigr]
 \;, 
 \la{initial_dvarphi_lat}
\end{align}
where 
$
 \tilde k \equiv |\tilde{\vec{k}}(\vec{n})| 
$
according to \eqs\nr{rescale2} and \nr{k_lat}; 
$\widetilde\varphi^{ }_k(0) \equiv {1} / {\sqrt{2 \widetilde \omega^{ }_k}}$ 
according to \eqs\nr{canonical} and \nr{rescale2}; 
and 
$
 \langle \, \hat c^{ }_\vec{n} \hat c^*_\vec{m} \, \rangle
 = 
 f^{ }_{k}\, \delta^{ }_{\vec{n},\vec{m}} 
$, 
where $f^{ }_k$ is from \eq\nr{variance_new}.

It is appropriate to stress that the mode function, 
$\widetilde\varphi^{ }_k$, involves the energy of the corresponding
mode, $\widetilde\omega^{ }_k$ (cf.\ \eq\nr{canonical}). On the lattice, 
because of the discretized Lagrangian in \eq\nr{eom_latt_scalar_resc}, 
the kinetic energy gets modified, becoming a trigonometric
function. However, we insert noise only 
in the IR domain $\tilde k \le 50 \ll 2\pi/\tilde a$
(cf.\ \eq\nr{def_tilde_a}), 
where the dispersion relation 
is to a reasonable approximation continuum-like. 

Power spectra at finite times are obtained from spatial Fourier
transforms, like in \eqs\nr{P_cl_pre} and \nr{P_cl}. On the lattice, 
and in rescaled units, the Fourier transform becomes
\ba
 G^{ }_{\widetilde\varphi}(t,\vec{k})
 & 
 \overset{\rmii{\nr{P_cl_pre}}}{
 \underset{\rmii{\nr{k_lat}}}{ = }} 
 & 
 \tilde a^3_{ } \sum_\vec{x} 
 e^{-i\frac{2\pi\vec{n}}{N}\cdot\frac{\vec{x}}{a}}_{ }
 \langle\,  
 \widetilde\varphi(t,\vec{x})
 \widetilde\varphi(t,\vec{0})
 \,\rangle
 \;, \quad
 \frac{\vec{x}}{a} 
 \;
 \in 
 \; 
 [0,1,...,N-1]^3_{ }
 \la{P_lat_pre}
 \\[3mm]
 & = & 
 \frac{1}{ \tilde a^3_{ }N^3_{ } }
 \bigl\langle\,  
 | \widetilde\varphi(t,\vec{k}) |^2_{ }
 \,\bigr\rangle
 \; = \; 
 \frac{ \tilde a^3_{ } }{ N^3_{ } }
 \biggl\langle\,  
 \biggl| 
    \sum_\vec{x} 
      e^{-i\frac{2\pi\vec{n}}{N}\cdot\frac{\vec{x}}{a}}_{ }
    \widetilde\varphi(t,\vec{x}) 
 \biggr|^2_{ }
 \,\biggr\rangle
 \;. 
 \la{fourier_norm}
\ea
Continuous 
rotational symmetry is absent,
but at small $\tilde k \ll 2\pi/\tilde a$ we can 
nevertheless define a power spectrum like in \eq\nr{P_cl}, 
\be
 \P^{ }_{\widetilde\varphi}(t,\tilde k)
 \; \equiv \; 
 \frac{\tilde k^3_{ } G^{ }_{\widetilde\varphi}(t,\vec{k})}{2\pi^2_{ }} 
 \;. \la{P_lat}
\ee

It is useful to carry out a consistency check
on the conventions adopted. If we take the initial condition from 
\eq\nr{initial_varphi_lat}, then 
\be
 \langle\,  
 \widetilde\varphi(0,\vec{x})
 \widetilde\varphi(0,\vec{0})
 \,\rangle
 \; 
 \overset{\rmii{\nr{initial_varphi_lat}}}{=}
 \; 
 \frac{1}{\tilde a^3_{ }N^3_{ }}
 \sum_{\vec{n}} |\widetilde\varphi^{ }_k(0)|^2_{ }
 \, f^{ }_k \, 
 \bigl[\,
    \, e^{i \vec{k}(\vec{n})\cdot \vec{x} }_{ }
 + 
    \, e^{-i \vec{k}(\vec{n})\cdot \vec{x} }_{ }
 \,\bigr]
 \;. \la{c_ck_1}
\ee
The Fourier transform then becomes 
(denoting its argument now by $\vec{q}$, 
and re-ordering sums)
\ba
 G^{ }_{\widetilde\varphi}(0,\vec{q})
 &
 \underset{\rmii{\nr{c_ck_1}}}{
 \overset{\rmii{\nr{P_lat_pre}}}{=}} 
 & 
 \frac{1}{N^3_{ }}
 \sum_{\vec{n}} |\widetilde\varphi^{ }_k(0)|^2_{ }
 \, f^{ }_k \, 
 \sum_{\vec{x}} 
 \bigl[\,
    \, e^{i ( \vec{k}(\vec{n}) -\vec{q} )\cdot \vec{x} }_{ }
 + 
    \, e^{-i ( \vec{k}(\vec{n}) + \vec{q} )\cdot \vec{x} }_{ }
 \,\bigr]
 \nn[2mm]
 & = & 
 \sum_{\vec{n}} |\widetilde\varphi^{ }_k(0)|^2_{ }
 \, f^{ }_k \, 
 \Bigl[\,
 \delta^{ }_{\vec{k}(\vec{n}),\vec{q}}
 + 
 \delta^{ }_{\vec{k}(\vec{n}),-\vec{q}}
 \,\Bigr]
 \nn[2mm]
 & = & 
 2 |\widetilde\varphi^{ }_q(0)|^2_{ }
 \, f^{ }_q
 \;.
\ea
Multiplying with the factor from \eq\nr{P_lat}, 
and inserting $f^{ }_q$ from \eq\nr{variance}, 
this reproduces the power spectrum in \eq\nr{def_P}
for small momenta, $q < \Lambda$. 

\vspace*{3mm}

Proceeding to the initial condition
for the electric field, we follow ref.~\cite{initialcond} and adopt a discretized 
version of \eq\nr{initial_Ei}. The spatial links 
are set to unity in the initial state, corresponding
to the vanishing of the magnetic field. 
The Gauss constraint from \eq\nr{Gauss_kappa1} then drops out.
In order to satisfy \eq\nr{Gauss_kappa0}, the polarization 
vectors need to fullfill
\be
 0 
 \;
 \underset{U^{ }_i \; =  \; \mathbbm{1} }{
 \overset{\rmii{\nr{Gauss_kappa0}}}{=}}
 \; 
 \sum_i (\vec{e}^\lambda_\vec{k})^{ }_i 
 \, \frac{2}{\tilde a} 
    \sin \biggl( \frac{\tilde a \tilde k^{ }_i}{2} \biggr)
 \; 
 \overset{\tilde a \; \ll \; 1}{\approx} 
 \;
 \sum_i (\vec{e}^\lambda_\vec{k})^{ }_i 
 \tilde k^{ }_i
 \;. \la{initial_Ei_latt}
\ee
Starting with an ansatz, 
for instance 
$\vec{e}_\vec{k}^0 \equiv (1,1,\sqrt{2})^\rmii{T}_{ }/2$, 
and denoting $\vec{v} \equiv \tilde{\vec{k}}/\tilde k$, 
two (linear) polarization 
states can be constructed as 
$
 \vec{e}_\vec{k}^1 
 = 
 \mathcal{N} \, 
 [
 \vec{e}_\vec{k}^0 - 
 (\vec{e}_\vec{k}^0\cdot \vec{v}) 
 \, \vec{v} 
 ]
$
and 
$
 \vec{e}_\vec{k}^2 = \vec{v} \times \vec{e}_\vec{k}^1
$, 
where $\mathcal{N}$ is a normalization factor. 
Given that the components of $\tilde{\vec{k}}$ are rational
(modulo $\pi$; cf.\ \eq\nr{k_lat}), 
the irrational $\sqrt{2}$ in $\vec{e}_\vec{k}^0$
guarantees that 
$ 
 \vec{e}_\vec{k}^1
 \neq
 \vec{0}
 \neq
 \vec{e}_\vec{k}^2
$ 
for all $\vec{k}$.


\newpage

\small{
%


\begin{thebibliography}{999}

\bibitem{planck6}
  N.~Aghanim {\it et al} [Planck],
  {\it Planck 2018 results.\ VI.\ Cosmological parameters,}
  \href{https://doi.org/10.1051/0004-6361/201833910}%
  {Astron.\ Astrophys.\ {641} (2020) A6}; 
  \href{https://doi.org/10.1051/0004-6361/201833910e}%
  {{\em ibid.} {652} (2021) C4 (erratum)} 
  [\href{https://arxiv.org/abs/1807.06209}{1807.06209}].

\bibitem{act}
  T.~Louis \textit{et al.}, 
  {\it The Atacama Cosmology Telescope: DR6 power spectra,
     likelihoods and $\Lambda$CDM parameters,}
     \href{https://doi.org/10.1088/1475-7516/2025/11/062}%
     {JCAP {11} (2025) 062}
  [\href{https://arxiv.org/abs/2503.14452}{2503.14452}].

\bibitem{spt}   
  J.A.~Zebrowski \textit{et al.} [SPT-3G],
  {\it SPT=3G D1: Constraints on inflationary gravitational waves with
  two years of SPT-3G data,}
  \href{https://doi.org/10.1103/33wm-c6pt}%
  {Phys.\ Rev.\ D {112} (2025) 123520}
  [\href{https://arxiv.org/abs/2505.02827}{2505.02827}].

\bibitem{ai}
  K.~Freese, J.A.~Frieman and A.V.~Olinto,
  {\it Natural inflation with pseudo Nambu-Goldstone bosons,}
  \href{https://doi.org/10.1103/PhysRevLett.65.3233}%
  {Phys.\ Rev.\ Lett.\  {65} (1990) 3233}.

\bibitem{cp1}
  R.D.~Peccei and H.R.~Quinn,
  {\it CP Conservation in the Presence of Pseudoparticles,}
  \href{https://doi.org/10.1103/PhysRevLett.38.1440}%
  {Phys.\ Rev.\ Lett.\ {38} (1977) 1440}.

\bibitem{cp2}
  S.~Weinberg,
  {\it A New Light Boson?,}
  \href{https://doi.org/10.1103/PhysRevLett.40.223}%
  {Phys.\ Rev.\ Lett.\ {40} (1978) 223}.

\bibitem{cp3}
  F.~Wilczek,
  {\it Problem of Strong  $P$  and  $T$  Invariance
  in the Presence of Instantons,}
  \href{https://doi.org/10.1103/PhysRevLett.40.279}%
  {Phys.\ Rev.\ Lett.\ {40} (1978) 279}.

\bibitem{heavy_ax}
  F.~Takahashi and W.~Yin,
  {\it Challenges for heavy QCD axion inflation,}
  \href{https://doi.org/10.1088/1475-7516/2021/10/057}%
  {JCAP {10} (2021) 057}
  [\href{https://arxiv.org/abs/2105.10493}{2105.10493}].

\bibitem{as}
  M.M.~Anber and L.~Sorbo,
  {\it Naturally inflating on steep potentials through
  electromagnetic dissipation,}
  \href{https://doi.org/10.1103/PhysRevD.81.043534}%
  {Phys.\ Rev.\ D {81} (2010) 043534}
  [\href{https://arxiv.org/abs/0908.4089}{0908.4089}].

\bibitem{as2}
  M.M.~Anber and L.~Sorbo,
  {\it Non-Gaussianities and chiral gravitational waves
  in natural steep inflation,}
  \href{https://doi.org/10.1103/PhysRevD.85.123537}%
  {Phys.\ Rev.\ D {85} (2012) 123537}
  [\href{https://arxiv.org/abs/1203.5849}{1203.5849}].

\bibitem{axab_0}
  P.~Adshead, J.T.~Giblin, M.~Pieroni and Z.J.~Weiner,
  {\it Constraining axion inflation with gravitational waves from preheating,}
  \href{https://doi.org/10.1103/PhysRevD.101.083534}%
  {Phys.\ Rev.\ D {101} (2020) 8}
  [\href{https://arxiv.org/abs/1909.12842}{1909.12842}].

\bibitem{axab_1}
  D.G.~Figueroa \textit{et al.}, 
  {\it Nonlinear dynamics of axion inflation: A detailed lattice study,}
  \href{https://doi.org/10.1103/PhysRevD.111.063545}%
  {Phys.\ Rev.\ D {111} (2025) 063545}
  [\href{https://arxiv.org/abs/2411.16368}{2411.16368}].

\bibitem{axab_2}
  O.~Iarygina, E.I.~Sfakianakis and A.~Brandenburg,
  {\it Schwinger effect in axion inflation on a lattice,}
  \href{https://arxiv.org/abs/2506.20538}{2506.20538}.

\bibitem{axab_3}
  D.~Jamieson, A.~Caravano and E.~Komatsu,
  {\it Primordial power spectrum and bispectrum from 
  lattice simulations of axion-U(1) inflation,}
  \href{https://doi.org/10.1103/3zml-71jd}%
  {Phys.\ Rev.\ D {112} (2025) 103531}
  [\href{https://arxiv.org/abs/2507.22285}{2507.22285}].

\bibitem{axab_4}
  R.~von Eckardstein,
  {\it GEFF: The Gradient Expansion Formalism Factory ---
  A tool for inflationary gauge-field production,}
  \href{https://arxiv.org/abs/2510.12644}{2510.12644}.

\bibitem{equil}
  Y.~Fu, J.~Ghiglieri, S.~Iqbal and A.~Kurkela,
  {\it Thermalization of non-Abelian gauge theories
  at next-to-leading order},
  \href{https://doi.org/10.1103/PhysRevD.105.054031}%
  {Phys.\ Rev.\ D {105} (2022) 054031}
  [\href{https://arxiv.org/abs/2110.01540}{2110.01540}].

\bibitem{fixed_pt}
  W.~DeRocco, P.W.~Graham and S.~Kalia,
  {\it Warming up cold inflation,}
  \href{https://doi.org/10.1088/1475-7516/2021/11/011}%
  {JCAP {11} (2021) 011}
  [\href{https://arxiv.org/abs/2107.07517}{2107.07517}].

\bibitem{therm}   
  T.~Fujita, K.~Mukaida and T.~Tsuji,
  {\it Reheating after axion inflation,}
  \href{https://doi.org/10.1088/1475-7516/2025/07/002}%
  {JCAP {07} (2025) 002}
  [\href{https://arxiv.org/abs/2503.01228}{2503.01228}].

\bibitem{hook}
  E.~Broadberry, A.~Hook and S.~Mondal,
  {\it Warm Inflation with Pseudo-scalar Couplings,}
  \href{https://arxiv.org/abs/2505.07943}{2505.07943}.

\bibitem{ema}
  S.~Bhattacharya, M.~Fasiello, A.~Papageorgiou and E.~Dimastrogiovanni,
  {\it On the prospects of thermalization of axion-SU(2) inflation,}
  \href{https://doi.org/10.1088/1475-7516/2025/10/080}%
  {JCAP {10} (2025) 080}
  [\href{https://arxiv.org/abs/2506.11853}{2506.11853}].
  
\bibitem{mms}
  L.~McLerran, E.~Mottola and M.E.~Shaposhnikov,
  {\it Sphalerons and axion dynamics in high-temperature QCD,}
  \href{https://doi.org/10.1103/PhysRevD.43.2027}%
  {Phys.\ Rev.\ D {43} (1991) 2027}.

\bibitem{warm}
  M.~Laine and S.~Procacci,
  {\it Minimal warm inflation with complete medium response,}
  \href{https://doi.org/10.1088/1475-7516/2021/06/031}%
  {JCAP {06} (2021) 031}
  [\href{https://arxiv.org/abs/2102.09913}{2102.09913}].

\bibitem{mt}
  G.D.~Moore and M.~Tassler,
  {\it The sphaleron rate in SU(N) gauge theory,}
  \href{https://doi.org/10.1007/JHEP02(2011)105}%
  {JHEP {02} (2011) 105}
  [\href{https://arxiv.org/abs/1011.1167}{1011.1167}].

\bibitem{clgt}
  M.~Laine, L.~Niemi, S.~Procacci and K.~Rummukainen,
  {\it Shape of the hot topological charge density spectral function,}
  \href{https://doi.org/10.1007/JHEP11(2022)126}%
  {JHEP {11} (2022) 126}
  [\href{https://arxiv.org/abs/2209.13804}{2209.13804}].

\bibitem{mwi}
  K.V.~Berghaus, P.W.~Graham and D.E.~Kaplan,
  {\it Minimal warm inflation,}
  \href{https://doi.org/10.1088/1475-7516/2020/03/034}%
  {JCAP {03} (2020) 034}; 
  \href{https://doi.org/10.1088/1475-7516/2023/10/E02}%
  {{\it ibid.} 10 (2023) E02 (erratum)}
  [\href{https://arxiv.org/abs/1910.07525}{1910.07525}].

\bibitem{linde}
  J.~Yokoyama and A.D.~Linde,
  {\it Is warm inflation possible?,}
  \href{https://doi.org/10.1103/PhysRevD.60.083509}%
  {Phys.\ Rev.\ D {60} (1999) 083509}
  [\href{https://arxiv.org/abs/hep-ph/9809409}{hep-ph/9809409}].

\bibitem{mehrdad}
  M.~Mirbabayi and A.~Gruzinov,
  {\it Shapes of non-Gaussianity in warm inflation,}
  \href{https://doi.org/10.1088/1475-7516/2023/02/012}%
  {JCAP {02} (2023) 012}
  [\href{https://arxiv.org/abs/2205.13227}{2205.13227}].

\bibitem{ballesteros}
  G.~Ballesteros, A.~P\'erez Rodr\'iguez and M.~Pierre,
  {\it Monomial warm inflation revisited,}
  \href{https://doi.org/10.1088/1475-7516/2024/03/003}%
  {JCAP {03} (2024) 003}
  [\href{https://arxiv.org/abs/2304.05978}{2304.05978}].

\bibitem{freese}
  G.~Montefalcone, V.~Aragam, L.~Visinelli and K.~Freese,
  {\it WarmSPy: a numerical study of cosmological perturbations 
  in warm inflation,}
  \href{https://doi.org/10.1088/1475-7516/2024/01/032}%
  {JCAP {01} (2024) 032}
  [\href{https://arxiv.org/abs/2306.16190}{2306.16190}].

\bibitem{ramos}
  G.S.~Rodrigues and R.O.~Ramos,
  {\it WI2easy: warm inflation dynamics made easy,}
  \href{https://doi.org/10.1088/1475-7516/2025/09/014}
  {JCAP {09} (2025) 014}
  [\href{https://arxiv.org/abs/2504.17760}{2504.17760}].

\bibitem{sm3}
  M.~Laine, S.~Procacci and A.~Rogelj,
  {\it Evolution of coupled scalar perturbations through
  smooth reheating.\ Part~II.\ Thermal fluctuation regime,}
  \href{https://doi.org/10.1088/1475-7516/2025/12/058}%
  {JCAP 12 (2025) 058}
  [\href{https://arxiv.org/abs/2507.12849}{2507.12849}].

\bibitem{sm1}
  K.V.~Berghaus, M.~Drewes and S.~Zell,
  {\it Warm Inflation with the Standard Model,}
  \href{https://doi.org/10.1103/9nn9-bsm9}%
  {Phys.\ Rev.\ Lett.\ 135 (2025) 171002}
  [\href{https://arxiv.org/abs/2503.18829}{2503.18829}].

\bibitem{sm2}
  R.O.~Ramos and G.S.~Rodrigues,
  {\it Viability of warm inflation with standard model interactions,}
  \href{https://doi.org/10.1103/wn1m-19gt}%
  {Phys.\ Rev.\ D {111} (2025) 123527}
  [\href{https://arxiv.org/abs/2504.20943}{2504.20943}].

\bibitem{Bastero-Gil:2014raa}
 M.~Bastero-Gil, A.~Berera, I.G.~Moss and R.O.~Ramos,
 {\it Theory of non-Gaussianity in warm inflation,}
 \href{https://doi.org/10.1088/1475-7516/2014/12/008}%
 {JCAP 12 {2014} 008}
 [\href{https://arxiv.org/abs/1408.4391}{1408.4391}].

\bibitem{mm}
  M.~Mirbabayi,
  {\it Loosely coupled particles in warm inflation,}
  \href{https://doi.org/10.1088/1475-7516/2025/05/067}%
  {JCAP {05} (2025) 067}
  [\href{https://arxiv.org/abs/2409.17927}{2409.17927}].

\bibitem{sorbo}
  Y.~Qiu and L.~Sorbo,
  {\it Spectrum of tensor perturbations in warm inflation,}
  \href{https://doi.org/10.1103/PhysRevD.104.083542}%
  {Phys.\ Rev.\ D {104} (2021) 083542}
  [\href{https://arxiv.org/abs/2107.09754}{2107.09754}].

\bibitem{reheat}
  P.~Klose, M.~Laine and S.~Procacci,
  {\it Gravitational wave background from non-Abelian
  reheating after axion-like inflation,}
  \href{https://doi.org/10.1088/1475-7516/2022/05/021}%
  {JCAP {05} (2022) 021}
  [\href{https://arxiv.org/abs/2201.02317}{2201.02317}].

\bibitem{klose}
  P.~Klose, M.~Laine and S.~Procacci,
  {\it Gravitational wave background from vacuum and
  thermal fluctuations during axion-like inflation,}
  \href{https://doi.org/10.1088/1475-7516/2022/12/020}%
  {JCAP {12} (2022) 020}
  [\href{https://arxiv.org/abs/2210.11710}{2210.11710}].

\bibitem{new}
  G.~Montefalcone, B.~Shams~Es~Haghi, T.~Xu and K.~Freese,
  {\it Thermal gravitons from warm inflation}, 
  \href{https://doi.org/10.1103/rnvb-t4lx}%
  {Phys.\ Rev.\ D {112} (2025) 063556}
  [\href{https://arxiv.org/abs/2507.08739}{2507.08739}].

\bibitem{new2}
  Q.~Chen \textit{et al.}, 
  {\it Freeze-in gravitational waves and dark matter in warm inflation,}
  \href{https://doi.org/10.1088/1475-7516/2026/03/051}%
  {JCAP {03} (2026) 051}
  [\href{https://arxiv.org/abs/2507.13916}{2507.13916}].

\bibitem{helena}
  H.~Kolesova, M.~Laine and S.~Procacci,
  {\it Maximal temperature of strongly-coupled dark sectors,}
  \href{https://doi.org/10.1007/JHEP05(2023)239}%
  {JHEP {05} (2023) 239}
  [\href{https://arxiv.org/abs/2303.17973}{2303.17973}].

\bibitem{Domcke:2019lxq}
 V.~Domcke and S.~Sandner,
 {\it The Different Regimes of Axion Gauge Field Inflation,}
 \href{https://doi.org/10.1088/1475-7516/2019/09/038}%
 {JCAP {09} (2019) 038}
 [\href{https://arxiv.org/abs/1905.11372}{1905.11372}].

\bibitem{pencil}
  O.~Iarygina, E.I.~Sfakianakis, R.~Sharma and A.~Brandenburg,
  {\it Backreaction of axion-SU(2) dynamics during inflation,}
  \href{https://doi.org/10.1088/1475-7516/2024/04/018}%
  {JCAP {04} (2024) 018}
  [\href{https://arxiv.org/abs/2311.07557}{2311.07557}].


\bibitem{bau}
  Z.-G.~Mou, P.M.~Saffin and A.~Tranberg,
  {\it Simulations of Cold Electroweak Baryogenesis: 
  hypercharge U(1) and the creation of helical magnetic fields,}
  \href{https://doi.org/10.1007/JHEP06(2017)075}%
  {JHEP {06} (2017) 075}
  [\href{https://arxiv.org/abs/1704.08888}{1704.08888}].

\bibitem{scalar}
  P.~Adshead, J.T.~Giblin and Z.J.~Weiner,
  {\it Non-Abelian gauge preheating,}
  \href{https://doi.org/PhysRevD.96.123512}%
  {Phys.\ Rev.\ D {96} (2017) 123512}
  [\href{https://arxiv.org/abs/1708.02944}{1708.02944}].


\bibitem{cl1}
  S.Y.~Khlebnikov and I.I.~Tkachev,
  {\it Classical decay of inflaton,}
  \href{https://doi.org/10.1103/PhysRevLett.77.219}%
  {Phys.\ Rev.\ Lett.\ {77} (1996) 219}
  [\href{https://arxiv.org/abs/hep-ph/9603378}{hep-ph/9603378}].

\bibitem{cl2}
  A.~Rajantie, P.M.~Saffin and E.J.~Copeland,
  {\it Electroweak preheating on a lattice,}
  \href{https://doi.org/10.1103/PhysRevD.63.123512}%
  {Phys.\ Rev.\ D {63} (2001) 123512}
  [\href{https://arxiv.org/abs/hep-ph/0012097}{hep-ph/0012097}].

\bibitem{chromo1}
  A.~Maleknejad and M.M.~Sheikh-Jabbari,
  {\it Gauge-flation: Inflation from non-Abelian gauge fields,}
  \href{https://doi.org/10.1016/j.physletb.2013.05.001}%
  {Phys.\ Lett.\ B {723} (2013) 224}
  [\href{https://arxiv.org/abs/1102.1513}{1102.1513}].

\bibitem{chromo2}
  P.~Adshead and M.~Wyman,
  {\it Natural Inflation on a Steep Potential 
  with Classical Non-Abelian Gauge Fields,}
  \href{https://doi.org/10.1103/PhysRevLett.108.261302}%
  {Phys.\ Rev.\ Lett.\ {108} (2012) 261302}
  [\href{https://arxiv.org/abs/1202.2366}{1202.2366}].


\bibitem{cosmo1}
  D.G.~Figueroa, A.~Florio, F.~Torrenti and W.~Valkenburg,
  {\it The art of simulating the early Universe. Part I. 
  Integration techniques and canonical cases,}
  \href{https://doi.org/10.1088/1475-7516/2021/04/035}%
  {JCAP {04} (2021) 035}
  [\href{https://arxiv.org/abs/2006.15122}{2006.15122}].

\bibitem{cosmo2}
  D.G.~Figueroa, A.~Florio, F.~Torrenti and W.~Valkenburg,
  {\it CosmoLattice: A modern code for lattice simulations of
  scalar and gauge field dynamics in an expanding universe,}
  \href{https://doi.org/10.1016/j.cpc.2022.108586}%
  {Comput.\ Phys.\ Commun.\ {283} (2023) 108586}
  [\href{https://arxiv.org/abs/2102.01031}{2102.01031}].

\bibitem{rk1}
  A.~Bazavov,
  {\it Commutator-free Lie group methods with minimum storage
  requirements and reuse of exponentials,}
  \href{https://doi.org/10.1007/s10543-021-00892-x}%
  {BIT Numerical Mathematics 62 (2022) 745}
  [\href{https://arxiv.org/abs/2007.04225}{2007.04225}].

\bibitem{rk2}
  A.~Bazavov and T.~Chuna,
  {\it Efficient integration of gradient flow in lattice gauge theory
  and properties of low-storage commutator-free Lie group methods,}
  \href{https://arxiv.org/abs/2101.05320}{2101.05320}.

\bibitem{old1}
  J.~Ambj{\o}rn, T.~Askgaard, H.~Porter and M.E.~Shaposhnikov,
  {\it Sphaleron transitions and baryon asymmetry:
  A numerical, real-time analysis,}
  \href{https://doi.org/10.1016/0550-3213(91)90341-T}%
  {Nucl.\ Phys.\ B {353} (1991) 346}.

\bibitem{cl3}
  G.D.~Moore,
  {\it Motion of Chern-Simons number at high temperatures
  under a chemical potential,}
  \href{https://doi.org/10.1016/S0550-3213(96)00445-2}%
  {Nucl.\ Phys.\ B {480} (1996) 657}
  [\href{https://arxiv.org/abs/hep-ph/9603384}{hep-ph/9603384}].

\bibitem{old}
  P.~Di Vecchia, K.~Fabricius, G.C.~Rossi and G.~Veneziano,
  {\it Preliminary evidence for U$^{ }_\rmii{A}$(1) breaking 
  in QCD from lattice calculations,}
  \href{https://doi.org/10.1016/0550-3213(81)90432-6}%
  {Nucl.\ Phys.\ B {192} (1981) 392}.

\bibitem{old2}
  I.T.~Drummond, S.~Duane and R.R.~Horgan,
  {\it The stochastic method for numerical simulations:
  Higher order corrections,}
  \href{https://doi.org/10.1016/0550-3213(83)90137-2}%
  {Nucl.\ Phys.\ B {220} (1983) 119}.

\bibitem{initialcond}
J.~Berges, K.~Boguslavski, S.~Schlichting and R.~Venugopalan,
{\it Universal attractor in a highly occupied non-Abelian plasma,}
\href{https://doi.org/10.1103/PhysRevD.89.114007}%
{Phys.\ Rev.\ D {89} (2014) 114007}
[\href{https://arxiv.org/abs/1311.3005}{1311.3005}].

\end{thebibliography}
\end{document}